\documentclass[a4paper,12pt]{article}
\usepackage{epsfig}
\usepackage{amssymb,subfigure}
\usepackage{amsfonts}
\usepackage{amsmath}
\usepackage{euscript}
\usepackage{verbatim}
\usepackage{latexsym}
\usepackage[compat=1.0.0]{tikz-feynman}
\usepackage{graphicx,color}
\usepackage{caption}
\usepackage{float}
\usepackage{slashed}
\usepackage{graphicx}
\usepackage{multirow}
\usepackage{afterpage}
\graphicspath{ {images/} }
\usepackage{ytableau}
\usepackage{mathtools}

\usepackage[colorinlistoftodos]{todonotes}
\usepackage[colorlinks=true, allcolors=blue]{hyperref}

\usepackage{wrapfig}
	\usepackage[T1]{fontenc}
	\usepackage{tikz}
	\usetikzlibrary{decorations.pathmorphing}
\usepackage{tikz}
\usetikzlibrary{shapes.geometric, arrows,patterns,snakes}
\tikzstyle{ellip} = [ellipse, minimum width=3cm, minimum height=1cm,text centered, draw=black]

\newskip\humongous \humongous=0pt plus 1000pt minus 1000pt

\newif\ifdtup

\allowdisplaybreaks[1]

\catcode`\@=11

\@addtoreset{equation}{section}

\def\@normalsize{\@setsize\normalsize{15pt}\xiipt\@xiipt
\abovedisplayskip 14pt plus3pt minus3pt%
\belowdisplayskip \abovedisplayskip
\abovedisplayshortskip \z@ plus3pt%
\belowdisplayshortskip 7pt plus3.5pt minus0pt}

\def\small{\@setsize\small{13.6pt}\xipt\@xipt
\abovedisplayskip 13pt plus3pt minus3pt%
\belowdisplayskip \abovedisplayskip
\abovedisplayshortskip \z@ plus3pt%
\belowdisplayshortskip 7pt plus3.5pt minus0pt
\def\@listi{\parsep 4.5pt plus 2pt minus 1pt
     \itemsep \parsep
     \topsep 9pt plus 3pt minus 3pt}}

\relax

\catcode`@=12

\catcode`\@=11

\def\section{\@startsection{section}{1}{\z@}{3.5ex plus 1ex minus
   .2ex}{2.3ex plus .2ex}{\large\bf}}

\def\SymBoxes#1#2#3#4{\newdimen\un@t \un@t#3%
\raisebox{#1}{\rule{#2\un@t}{#4}\hskip-#2\un@t
\@tempdimb\un@t \advance\@tempdimb by-#4\@tempcntb#2\relax%
\@whilenum{\@tempcntb>0}\do{
\rule{#4}{\un@t}\hskip\@tempdimb \advance\@tempcntb by\m@ne}%
\hskip-#2\un@t \rule[\un@t]{#2\un@t}{#4}%
\rule[\un@t]{#4}{#4}\hskip-#4
\rule{#4}{\un@t}}\hskip-#4}                

\DeclareMathAlphabet{\pazocal}{OMS}{zplm}{m}{n}

\begin{document}

\newcommand{\beq}{\begin{equation}}
\newcommand{\eeq}{\end{equation}}
\newcommand{\bea}{\begin{eqnarray}}
\newcommand{\eea}{\end{eqnarray}}
\newcommand{\beas}{\begin{eqnarray*}}
\newcommand{\eeas}{\end{eqnarray*}}
\newcommand{\defi}{\stackrel{\rm def}{=}}
\newcommand{\non}{\nonumber}
\newcommand{\bquo}{\begin{quote}}
\newcommand{\enqu}{\end{quote}}
\renewcommand{\(}{\begin{equation}}
\renewcommand{\)}{\end{equation}}
\def \eqn#1#2{\begin{equation}#2\label{#1}\end{equation}}
\def\IZ{{\mathbb Z}}
\def\IR{{\mathbb R}}
\def\IC{{\mathbb C}}
\def\IQ{{\mathbb Q}}
\def\de{\partial}
\def\Tr{ \hbox{\rm Tr}}
\def\H{ \hbox{\rm H}}
\def\HE{ \hbox{$\rm H^{even}$}}
\def\HO{ \hbox{$\rm H^{odd}$}}
\def\K{ \hbox{\rm K}}
\def\Im{ \hbox{\rm Im}}
\def\Ker{ \hbox{\rm Ker}}
\def\const{\hbox {\rm const.}}
\def\o{\over}
\def\im{\hbox{\rm Im}}
\def\re{\hbox{\rm Re}}
\def\bra{\langle}\def\ket{\rangle}
\def\Arg{\hbox {\rm Arg}}
\def\Re{\hbox {\rm Re}}
\def\Im{\hbox {\rm Im}}
\def\exo{\hbox {\rm exp}}
\def\diag{\hbox{\rm diag}}
\def\longvert{{\rule[-2mm]{0.1mm}{7mm}}\,}
\def\a{\alpha}
\def\dag{{}^{\dagger}}
\def\tq{{\widetilde q}}
\def\p{{}^{\prime}}
\def\W{W}
\def\N{{\cal N}}
\def\hsp{,\hspace{.7cm}}

\def\br{\nonumber\\}
\def\IZ{{\mathbb Z}}
\def\IR{{\mathbb R}}
\def\IC{{\mathbb C}}
\def\IQ{{\mathbb Q}}
\def\IP{{\mathbb P}}
\def \eqn#1#2{\begin{equation}#2\label{#1}\end{equation}}
\def\t{\tilde}

\newcommand{\sgm}[1]{\sigma_{#1}}
\newcommand{\idd}{\mathbf{1}}

\newcommand{\C}{\ensuremath{\mathbb C}}
\newcommand{\Z}{\ensuremath{\mathbb Z}}
\newcommand{\R}{\ensuremath{\mathbb R}}
\newcommand{\rp}{\ensuremath{\mathbb {RP}}}
\newcommand{\cp}{\ensuremath{\mathbb {CP}}}
\newcommand{\vac}{\ensuremath{|0\rangle}}
\newcommand{\vact}{\ensuremath{|00\rangle}                    }
\newcommand{\oc}{\ensuremath{\overline{c}}}
\begin{titlepage}
\begin{flushright}
CHEP XXXXX
\end{flushright}
\bigskip
\def\thefootnote{\fnsymbol{footnote}}

\begin{center}
{\large
{\bf Celestial Klein Spaces 
}
}
\end{center}

\bigskip
\begin{center}
{Budhaditya BHATTACHARJEE$^a$\footnote{\texttt{budhaditya95@gmail.com}}, \ \ \ Chethan KRISHNAN$^a$\footnote{\texttt{chethan.krishnan@gmail.com}} }
\vspace{0.1in}


\end{center}

\renewcommand{\thefootnote}{\arabic{footnote}}

\begin{center}
$^a$ {Center for High Energy Physics,\\
Indian Institute of Science, Bangalore 560012, India}

\end{center}

\begin{center} {\bf Abstract} \end{center}
We consider the analytic continuation of $(p+q)$-dimensional Minkowski space (with $p$ and $q$ even) to $(p,q)$-signature, and study the conformal boundary of the resulting ``Klein space''. Unlike the familiar $(-+++..)$ signature, now the null infinity ${\cal I}$ has only one connected component. The spatial and timelike infinities ($i^0$ and $i'$) are quotients of generalizations of AdS spaces to non-standard signature. Together, ${\cal I}, i^0$ and $i'$ combine to produce the topological boundary $S^{p+q-1}$ as an $S^{p-1} \times S^{q-1}$ fibration over a null segment. The highest weight states (the $L$-primaries) and descendants of $SO(p,q)$ with integral weights give rise to natural scattering states. One can also define $H$-primaries which are highest weight with respect to a signature-mixing version of the Cartan-Weyl generators that leave a point on the celestial $S^{p-1} \times S^{q-1}$ fixed. These correspond to massless particles that emerge at that point and are Mellin transforms of plane wave states.  



\vspace{1.6 cm}
\vfill

\end{titlepage}

\setcounter{page}{2}

\setcounter{footnote}{0}

\section{Introduction}

Analytic continuation in momentum space has proven to be a powerful tool for understanding scattering amplitudes (see \cite{Eden} for a historical reference), so it is natural to suspect that generalizations of the S-matrix from 3+1 Minkowski space to other signatures may also provide further insights on scattering. A key reason to suspect this is that the on-shell structure of scattering amplitudes is largely controlled by the 3-point amplitude (see, e.g., \cite{Elvang}), but the 3-point amplitude identically vanishes in 3+1 (or, for that matter, Euclidean) signature due to kinematics. Considering the fact that on-shell methods have been powerful tools in the last decade or so in developing scattering amplitude technology, this is a natural motivation to consider signature with some ``fluidity".
\\

An interesting step in this direction was recently taken in \cite{Strominger}, where analytic continuation of Minkowski space to (2,2) signature was considered. It was found that, unlike in the (3,1) signature case, the conformal boundary has the form of a celestial torus, and suitable scattering states associated with an ``S-vector'' (as opposed to an S-matrix) were constructed. This work is in the broader context of celestial holography; see \cite{summary} for points of entry into the recent literature. In this paper, we will generalize the results of \cite{Strominger} to $p+q$ dimensions, with signature $(p,q)$ where $p$ and $q$ are even.
\\

While our general motivations for considering higher dimensions should be obvious to string theorists, let us take a moment to note a few points which may be less obvious. Firstly, the argument made above about the on-shell 3-point amplitude is a kinematic restriction. Signature fluidity is more crucial for kinematics than the demand that it necessarily be (2,2). So in higher dimensions, it makes sense to consider $(p,q)$ signature. A second argument arises from holography. The conformal boundary of Euclidean flat space is a point, but as we will demonstrate in any $(p,q)$ signature, the boundary has more structure. So we would like to avoid being tied to split signature. Yet another thing to note is that (2,2) signature is closely connected to self-duality conditions, which is a dynamical restriction. We wish to leave the door open for questions that are likely to be more generic. 
\\

One of the technical features of four dimensions is that the Lorentz group has simplifying features. This allows us to exploit those simplifications, and treat four dimensions using special methods. In quantum field theory classes, this manifests itself in the observation that we can work with left handed and right handed Weyl spinors instead of Dirac spinors in four dimensions. A morally similar phenomenon occurs in (2,2) signature, where the isometry group is $SO(2,2)$. Because of its connections to the global conformal group $SL(2,R)\times SL(2,R)$ in 2 dimensions, this resulted in a discussion \cite{Strominger} that was somewhat special. In particular, the scattering states were associated with $SL(2,R)$ highest weight states. In higher dimensions, we do not have the benefit of such accidental Lie algebra isomorphisms or factorization of the algebra. 
\\

Despite this, in this paper we will show that the discussion of \cite{Strominger} is, in fact, much more general and that both the geometry and the group theory generalize very naturally to arbitrary even dimensions. We do this by taking advantage of the Cartan-Weyl form of the Lie algebra in higher dimensions. We consider general $(p+q)$-dimensional Minkowski space with $p$ and $q$ even, and its analytic continuation to $(p,q)$ signature. Following the terminology of \cite{Strominger}, we will call these spaces Klein spaces. Instead of the celestial sphere, at the conformal boundary, we find a celestial $S^{p-1}\times S^{q-1}$ space. Even though we are in higher dimensions, we find that we can construct suitable scattering states (the $L$-primaries) associated with the highest weight states, which we explicitly construct starting with the Cartan-Weyl-like form of the $SO(p,q)$ algebra and a suitable foliation of the Klein space. We can also construct a similar class of states (the $H$-primaries) which are defined via Cartan generators that mix the signatures, in such a way that the transformations fix a point on the celestial  $S^{p-1}\times S^{q-1}$. These states can be written as Mellin transforms of plane waves. The higher dimensional picture makes it clear that the two classes of primaries are naturally thought of as rotation-like and boost-like.
\\

The paper is structured as follows. In the Section II, we discuss the
geometric structure of pseudo-Euclidean spaces with $SO(p,q)$
isometry, which we call Klein spaces $\mathbf{K}^{p,q}$. We also
define the coordinates and notation that we use for the rest of the
paper. In Section III, we describe the algebraic structure of the
$SO(p,q)$ group and write down the Cartan-Weyl form
(``raising-lowering form'') of the algebra. We also write down the generators as differential operators in
Klein space coordinates. Because of the somewhat elaborate nature of the index structures involved, a summary of index notation is provided. We write
down the Casimir of the group and discuss the representations of the
group in terms of the associated eigenfunctions. In Section IV, we
explicitly solve the wave equation in the Klein space (we call these
$L-$primaries) and derive the solutions that correspond to specific
weights (highest/lowest/mixed) with respect to the Casimir. In Section
V, we consider a different set of raising-lowering operators (a
different basis) where some of the Cartan generators are boost-type.
The states corresponding to this choice are particles that emerge at
some point on the celestial $S^{p-1}\times S^{q-1}$. We repeat the same
procedure as before for the states corresponding to various weights in
this basis (called $H-$primaries). Finally, we relate the $H-$ and
$L-$primaries (and their descendants) by demonstrating that one can
be written as an integral transform of the other. We also note that
the $H-$primary states can be written as Mellin transforms of plane
waves \cite{Strominger}. The concluding section of the main text contains a comparison between our notations and those of \cite{Strominger}, estabilishing the precise match when $p=q=2$. In Appendix A-C we explicitly write down all
generators in terms of the coordinates in the Klein space and show the
steps involved in solving for the primaries. In Appendix D, we briefly
review some facts about the quadratic Casimir.
\\

Our paper should be viewed as an extension of the results in \cite{Strominger}. So while we have made an effort to be technically self-contained, we encourage the reader to consult \cite{Strominger} and references therein, for more background and motivations. 
\section{Klein Spaces}

\subsection{Defining Coordinates}\label{sec2.1}

We will consider pseudo-Euclidean spaces with an $SO(p,q)$ isometry, where $p = 2 n$ and $q = 2 m$ and $n$ and $m$ are positive, non-zero integers. We will call such a $(2n,2m)$ signature ``space-time'', a flat Klein space $\mathbf{K}^{2n,2m}$. Some of our discussions, in fact, apply more generally to the case when the even-ness condition on $p$ and $q$ is relaxed.

The metric of such a space is 
\begin{equation}\label{1.2}
ds^{2} = - \sum_{a=1}^{2n}dx_{a}^{2} + \sum_{\tilde{a}=1}^{2m}dy_{\tilde{a}}^{2} .
\end{equation}
We introduce complex coordinates
\begin{eqnarray}
z_{i} &= x_{2i-1} + i x_{2 i}\label{1.3} \\
w_{\t{i}} &= y_{2\t{i}-1} + i y_{2\t{i}}\\
\bar{z}_{i} &= x_{2i-1} - i x_{2 i} \\
\bar{w}_{\t{i}} &= y_{2\t{i}-1} - i y_{2\t{i}}\label{1.6}
\end{eqnarray}
where $i$ goes from $1$ to $n$ and $\t{i}$ goes from $1$ to $m$. Note that the indices in \eqref{2.1} $a, \tilde{a}$ run over double the range of the $i,\t{i}$ coordinates. We have the following expression for the metric in terms of the complex coordinates 
\begin{equation}\label{1.8}
ds^{2} = \sum_{\t{j}=1}^{m}dw_{\t{j}}d\bar{w}_{\t{j}} - \sum_{j=1}^{n}dz_{j}d\bar{z}_{j}
\end{equation}
With future use in mind, let us also parametrize these complex variables as 
\begin{equation}
z_{i} = r_{i}e^{i \theta_{i}}, \; \;
w_{\t{i}} = t_{\t{i}}e^{i \phi_{\t{i}}} \label{1.10}
\end{equation}
and we have
\begin{equation}\label{1.12}
ds^{2} = - \sum_{i=1}^{n}(dr_{i}^{2} + r_{i}^{2}d\theta_{i}^{2}) + \sum_{\t{i}=1}^{m}(dt_{\t{i}}^{2} + t_{\t{i}}^{2}d\phi_{\t{i}}^{2})
\end{equation}
This last parametrization will turn out to be very useful for our purposes. These cordinates were not needed in \cite{Strominger}, because one could exploit the connection between $SO(2,2)$ and $SL(2,R) \times SL(2,R)$ instead. 



\subsection{Null, Space-like and Time-like Infinity}

We will conformally compactify the $\mathbf{K}^{p,q}$ space and determine the geometry of null infinity $\mathcal{I}$, spatial infinity $i^{0}$ and time-like infinity $i'$. Writing the line element as
\begin{equation}\label{2.1}
ds^{2} = -\sum_{a=1}^{p}dx_{a}^{2} + \sum_{\tilde{a}=1}^{q}dy_{\tilde{a}}^{2} 
\end{equation}
and introducing spherical polar coordinates with radii $r$ and $t$ for $x_{a}$ and $y_{\t{a}}$, we get 
\begin{equation}\label{2.2}
ds^{2} = -dr^{2} - r^{2}d\Omega^{2}_{p-1} + dt^{2} + t^{2}d\tilde{\Omega}_{q-1}^{2}
\end{equation}
We can now shift to a ``lightcone'' coordinate system via
\bea\label{2.3}
r-t = \tan U, \ \ 
r+t = \tan V 
\eea
Making this substitution in the Eqn \eqref{2.2}, we get
\begin{equation}\label{2.5}
ds^{2} = \frac{1}{\cos U^{2} \cos V^{2}}(-dU dV -\frac{1}{4}\sin^{2}(V+U)d\Omega^{2}_{p-1} +\frac{1}{4}\sin^{2}(V-U)d\tilde{\Omega}^{2}_{q-1} )
\end{equation}
\par
Null infinity $\mathcal{I}$ is at $V = \frac{\pi}{2}$ where the factor out front blows up. Space-like (time-like) infinity $i_{0}$ ($i'$) is the boundary at $U = -\frac{\pi}{2}$ ($U= \frac{\pi}{2}$).
The surface at $\mathcal{I}$ is given by $-\frac{\pi}{2} < U < \frac{\pi}{2}$. We take the limit $V \rightarrow \frac{\pi}{2}$ and scale out the $\cos^{2}V$ term. Then we get
\begin{equation}\label{2.6}
ds^{2}_{\mathcal{I}} = -d\Omega^{2}_{p-1} + d\tilde{\Omega}^{2}_{q-1}
\end{equation}
where the angles in $\Omega$ and $\tilde{\Omega}$ have the periodicities and ranges of the corresponding spheres. The spacelike sphere degenerates along the timelike line $U=V$ and the timelike sphere degenerates along the spacelike line $U=-V$. Neither sphere degenerate at the null infinity $\mathcal{I}$.

The topological boundary of $\mathbf{K}^{p,q}$ is $S^{p+q-1}$ and we can view it as an $S^p \times S^q$ fibration over a segment. To establish that ${\cal I}$ together with $i_0$ and $i'$ has this topology, we need to show that  each of the ends $i_0$ and $i'$ have the topology of a ball times a sphere (of appropriate dimensions).  \par
To see this, let us foliate the Klein space using the following slicing:
\begin{equation}\label{2.7}
\sum_{\tilde{a}=1}^{q}y_{\tilde{a}}^{2} - \sum_{a=1}^{p}x_{a}^{2} = \pm \tau^{2}
\end{equation}
The two regions corresponding to the $\pm$ sign of $\tau^{2}$ can be denoted by $\mathbf{K}^{p,q \pm}$. Note that it is the region $\mathbf{K}^{p,q -}$, which contains the time-like infinity. 
So we start by considering $-\tau^{2}$ case. The goal is to then take $\tau \rightarrow \infty$ to end up at the ``upper half'' of $\mathcal{I}$. 
We can parametrize $x$ and $y$ in spherical coordinates, as Eqn \eqref{2.2}, but with
\begin{align}\label{2.8to9}
r &= \tau \cosh \rho \\
t &= \tau \sinh \rho 
\end{align}
This gives us
\begin{equation}\label{2.10}
ds^{2} = -d\tau^{2} + \tau^{2}ds_{1}^{2}
\end{equation}
\begin{equation}\label{2.11pan}
ds_{1}^{2} = d\rho^{2} -\cosh^{2}\rho\; d\Omega^{2}_{p-1} + \sinh^{2}\rho\; d\tilde{\Omega}^{2}_{q-1}
\end{equation}
The metric \eqref{2.11pan} is known as $AdS_{q,p-1}$, see eg.  \cite{wikipedia}. It contains $p-1$ time-like coordinates and $q$ space-like coordinates. From \eqref{2.10}, one notes that we are foliating the generalized Minkowski space $M_{q,p}$ using these generalized AdS spaces. 
\par
From this metric, we can look at the geometry of $i'$ by taking the limit of $\rho \rightarrow 0$. One can see this by noting that going to $i'$ means that we need to shrink the ``space-like radius $t$'' to a small value. This leads to a metric of the form
\begin{equation}\label{2.11}
	ds_{1}^{2} \sim d\rho^{2}  + \rho^{2}\; d\tilde{\Omega}^{2}_{q-1}- d\Omega^{2}_{p-1}
\end{equation}
There is a clear factorization in this geometry, which leads us to observe that the space-time becomes a topological product of the $q$-ball $\mathcal{S}^{q}$ and the $(p-1)$-sphere $S^{p-1}$. This is the higher dimensional generalization of the topological product of a circle and a disk that is obtained as the geometry of $i'$ in the $\mathbf{K}^{2,2 -}$ case \cite{Strominger}.\par
A similar construction holds for the case of $+\tilde{\tau}^{2}$. Here, the $\tilde{\tau} \rightarrow \infty$ leads us to the lower half of the null infinity $\mathcal{I}$. 
In this case, we use the alternative parametrization
\begin{eqnarray}\label{2.12to13}
r &= \tilde{\tau} \sinh \tilde{\rho}\\
t &= \tilde{\tau} \cosh \tilde{\rho}
\end{eqnarray}
Which gives us
\begin{equation}\label{2.14}
ds^{2} = d\tilde{\tau}^{2} -\tilde{\tau}^{2}ds_{1}^{2}
\end{equation}
\begin{equation}\label{2.15}
ds_{1}^{2} = d\tilde{\rho}^{2} - \cosh^{2}\tilde{\rho}\; d\tilde{\Omega}_{q-1}^{2} + \sinh^{2}\tilde{\rho}\;d\Omega_{p-1}^{2}
\end{equation}
This metric is the one known as $AdS_{p,q-1}$. 
We can take the $\tilde{\rho} \rightarrow 0$ limit to reach the spacelike infinity $i^{0}$, which corresponds to taking the  ``timelike radial coordinate $r$'' to a very small value. 
In this limit, the metric \eqref{2.15} again factorises into the following simple form
\begin{equation}\label{2.15}
	ds_{1}^{2} \sim d\tilde{\rho}^{2}  + \tilde{\rho}^{2}\;d\Omega_{p-1}^{2}-  d\tilde{\Omega}_{q-1}^{2}
\end{equation}
This leads us to observe that the geometry at $i^{0}$ is a topological product of a $p$-ball $\mathcal{S}^{p}$ and a $(q-1)$-sphere $S^{q-1}$. Therefore, we conclude that the geometry of $\mathcal{I}$ along with $i'$ and $i^{0}$ can be viewed as an $S^{p+q-1}$, where $S^{p-1}\times S^{q-1}$ is fibered over a line segment.
\par 
One point to note about our generalized $AdS$ foliation is that it is impossible to ``unwrap'' the timelike directions on these AdS spaces.  This is possible only in $AdS$ metrics that arise from an embedding space with only two time-like coordinates. In the present case, we have a sphere-worth of timelike coordinates instead of a circle. 
\\

\section{Klein, Cartan-Weyl and Casimir}

Highest weight states will play an important role in our discussion of the scattering states, and to discuss them, it is useful to work with the Cartan-Weyl basis of $SO(2n,2m)$. In the Cartan-Weyl basis, all the generators can be written in terms of the commutators of the \textit{Cartan generators}, which form an Abelian sub-algebra, and the \textit{ladder operators}. Since we are interested in only the even-dimensional cases, we will work with the complex Lie algebra $D_N$, which has $N$ Cartan generators. The basis is conveniently spanned by the eigenvectors of the Cartan generators. The ladder operators then raise or lower the eigenvalues for these states and enable us to conveniently define the highest weight states. 


To orient ourselves, let us start by writing down the general complex Lie algebra in the Cartan-Weyl basis. We have the Cartan generators $H_{i}$ and the raising and lowering operators corresponding to the \textit{roots} $\alpha$ of the algebra given by $E_{\pm \alpha}$. The commutation relations that they must satisfy are the following
\begin{align}\label{3.1}
[H_{i},H_{j}] &= 0\\
[H_{i},E_{\alpha}] &= \alpha^{i}E_{\alpha}\label{3.2}\\
\begin{split}
[E_{\alpha},E_{\beta}] &= N_{\alpha + \beta}E_{\alpha + \beta}\;\;\;\;\;\;\; \alpha + \beta\; \in \; \text{root}\\  
&= \frac{2}{|\alpha|^{2}}\alpha.H \;\;\;\;\;\;\;\;\;\; \alpha = -\beta\\
&= 0 \;\;\;\;\;\;\;\;\;\;\;\;\;\;\;\;\;\;\;\;\; \alpha + \beta \notin \;\text{root}\label{3.3}
\end{split}
\end{align}
The rest of the generators are encoded indirectly in these generators, and we will not need them. The above form corresponds to the complex Lie algebra, where one assumes that the vector space spanned by these generators is defined over the complex number field. A complex Lie algebra has many {\em real forms}. If one simply dictates that the field over which the above algebra is defined is the real numbers as opposed to the complex numbers, what one gets is called the {\em split real form} which in the case of $D_N$ corresponds to $SO(N,N)$. There is also a unique {\em compact real form} which in the case of $D_N$ is to be viewed as the Lie algebra of $SO(2N)$. We will be interested in the general {\em non-compact real forms}  $SO(p,q)$ of the complex Lie algebra $D_{(p+q)/2}$. 

We are interested in viewing scattering states as highest weight states, and it is for this reason that we will be interested in the Cartan-Weyl form of $SO(p,q)$. But the differential operator realization is most conveniently identified in the coordinates that manifest the Klein space isometries. Defining and relating these two realizations of the algebra will be the goal of this section.


\subsection{Klein Space Generators}

We first write down the algebra of the $SO(p,q)$ group as the isometry algebra of the underlying Klein space. Denoting generators by $J_{A B}$, the commutation relation that they  satisfy is 
\begin{equation}\label{comm}
	[ J_{A B}, J_{C D}] = i(\eta_{A D}J_{ B C} + \eta_{B C}J_{A D} - \eta_{A C}J_{B D} - \eta_{B D}J_{A C}) 
\end{equation}
where $\eta_{A B}$ is the flat Klein space metric with the components $(-\mathbf{1}_{p},\mathbf{1}_{q})$.  The standard Hermitian matrix form for the generators can now be written as 
\begin{equation}\label{3.51}
	[J_{A B}]^{\mu}_{\nu} = i \Big( \delta^{\mu}_{A}\eta_{\nu B} - \delta^{\mu}_{B}\eta_{\nu A}\Big)
\end{equation}
The index placement (upper or lower) of $\mu$ and $\nu$ not distinguished here, and they should be summed when repeated. 

We wish to give a differential operator realization of these generators using the underlying Klein space coordinates. 
\begin{align}
	X_\mu &\equiv \{ \{x_{a}\}, \{y_{\tilde{a}}\}  \} =  \{ x_{1},\cdots, x_{p},y_{1}, \cdots, y_{q} \}\label{coord}\\
	(\partial_{X})^{\nu} &\equiv \{ \{\partial_{x_{a}}\}, \{\partial_{y_{\tilde{a}}}\}  \} = \{\partial_{x_{1}}, \cdots, \partial_{x_{p}},\partial_{y_{1}}, \, \cdots, \partial_{y_{q}}\}
\end{align}
where we have used the abbreviated notation $\partial_{z} \equiv \frac{\partial}{\partial z}$, for any variable $z$. The $x$ coordinates are taken to be the ones corresponding to $-1$ signature and $y$ coordinates are taken to correspond to $+1$ signature, and follows the notation of section 2. 
 
Using these, we can write explicit differential operator expressions for $J_{A B}$ in terms of the coordinates as follows (we denote it by $M_{A B}$ to distinguish it from the matrix form $[J_{A B}]^{\mu}_{\nu}$ of the generators)
\begin{equation}\label{mab}
	M_{A B} \equiv  \sum_{\mu,\nu} X_{\mu}[J_{A B}]^{\mu}_{\nu}(\partial_{X})^{\nu}
\end{equation}
Evaluating this expression gives us the following results
\begin{align}
	M_{a b} &= -i \Big( x_{a}\partial_{x_{b}} - x_{b}\partial_{x_{a}}\Big)\label{gen1}  \\
	M_{\tilde{a} \tilde{b}} &= i(y_{\tilde{a}}\partial_{y_{\tilde{b}}} - y_{\tilde{b}}\partial_{y_{\tilde{a}}})\\
	M_{a \tilde{a}} &= i(x_{a}\partial_{y_{\tilde{a}}} + y_{\tilde{a}}\partial_{x_{a}})\label{gen3}
\end{align}
Here, the notation is as it was in section 2,
\begin{align*}
	a, b &\in 1,\dots,p \\
	\tilde{a}, \tilde{b} &\in 1, \dots, q
\end{align*}
with the un-tilded small case letters from early part of the Latin alphabet going with the $x$-coordinates and the tiled ones with the $y$-coordinates.
By noting that $\eta_{a b} = -\delta_{a b}$, $\eta_{\tilde{a} \tilde{b}} = \delta_{\tilde{a} \tilde{b}}$ and $\eta_{a \tilde{a}} = 0$, we can explicitly demonstrate that \eqref{gen1}-\eqref{gen3} exactly satisfies \eqref{comm}.

\subsection{Klein to Cartan-Weyl}

We now want to write down a Cartan-Weyl-like basis for the group $SO(2n,2m)$ and relate it to the differential generators of the last subsection. This will be useful in interpreting the Casimir as a Laplace-like operator and the raising/lowering/highest-weight/lowest-weight conditions in terms of differential operators. 

We will adapt the construction of our Cartan-Weyl form from the complex Lie algebra $D_{n+m}$, whose roots are explicitly given in terms of a Euclidean coordinate basis in \cite{Georgi} (see sections 19.1 and 21.1\footnote{The discussion in section 21.1 of \cite{Georgi} is for $B_n$ algebras, but it is easy enough to adapt to the present discussion.}). For the complex $D_{n+m}$, the roots of the algebra can be written in the form $\alpha_{J,K} \equiv \xi e^{J} + \xi' e^{K}$, where we have introduced new indices $J,K \; \in \; \{1,\dots\dots,n+m\}$ and $\xi , \xi'$ can take the values $\pm 1$. See \cite{Georgi} for the definition of $e^{J}$; we will not need them explicitly. The {\em positive roots}, which will be important for us in defining the highest weight states, are defined with $\xi = 1$, $\xi' = \pm 1$ and $K>J$. In the rest of the text, whenever we use $\alpha_{J,K}$, we will mean the positive roots only (and will therefore have $K>J$). We will introduce explicit negative signs when we need the negative roots. We will write the Cartan-Weyl form of the algebra in a form that splits the ladder operators in terms of positive and negative roots. 

To construct the complex form of the Lie algebra in the Cartan-Weyl form, one takes \cite{Georgi} the Cartan generators to be 
\begin{equation}
	H_{J} = M_{2J-1,2J}
\end{equation}  
and the ladder operators corresponding to \emph{any root} $\xi e^{J} + \xi' e^{K}$  to be 
\begin{equation}
	E_{\xi e^{J} + \xi' e^{K}} = \frac{1}{2}\Big( M_{2J-1,2K-1} + i\xi M_{2J,2K-1} + i\xi' M_{2J-1,2K} - \xi \xi' M_{2J,2K}	\Big). \label{Ladder-def}
\end{equation}
Here $M$ are taken to be the Euclidean generators of the compact real form of the algebra. 

In order to come up with a useful Cartan-Weyl form for $SO(2n,2m)$,  we will use the same expressions above, but now the $M$ generators will be viewed as those in the previous subsection - those that capture the Klein space isometry. This way, we will be able to relate the notion of a highest weight state, which is manifest in the Cartan-Weyl form, to the derivative operators that capture the Klein space isometry. In an appendix, we will show that the highest weight states that follow from our definition match precisely with the $SL(2,R)\times SL(2,R)$ construction of \cite{Strominger} when $p=q=2$.




The specific expressions for the Cartan generators and the Ladder operators can then be taken as follows. The Cartan generators are
\begin{equation}\label{Klein-Cartan}
	H_{J} = M_{2J-1,2J}.
\end{equation}
The Ladder operators corresponding to the positive roots are
\begin{align}\label{KleinRaising1}
	E_{e^{J} + e^{K}} &= \frac{1}{2}\{M_{2J-1,2K-1} + i M_{2J, 2K-1} + i(M_{2J-1,2K} + i M_{2J,2K})\}\\
	E_{e^{J}-e^{K}} &= \frac{1}{2}\{M_{2J-1,2K-1} + i M_{2J, 2K-1} - i(M_{2J-1,2K} + i M_{2J,2K})\}\label{KleinRaising2}
\end{align}
which we will collectively call $E_{\alpha_{J,K}}$ and those corresponding to the negative roots are  
\begin{align}\label{KleinLowering1}
	F_{e^{J} + e^{K}} &= \frac{1}{2}\{M_{2J-1,2K-1} - i M_{2J, 2K-1} - i(M_{2J-1,2K} - i M_{2J,2K})\}\\
	F_{e^{J}-e^{K}} &= \frac{1}{2}\{M_{2J-1,2K-1} - i M_{2J, 2K-1} + i(M_{2J-1,2K} - i M_{2J,2K})\}\label{KleinLowering2}
\end{align} 
which we will collectively call $E_{-\alpha_{J,K}}$.
The commutation relations satisfied by these operators are as follows
\begin{align}
	[H_{I},H_{J}] &= 0 \label{axx3}\\ 
	[H_{I},E_{e^{J} + e^{K}}] &= -(\eta_{2I, 2J} + \eta_{2I, 2K})E_{e^{J} + e^{K}} \label{bxx}\\ 
	[H_{I},E_{e^{J} - e^{K}}] &= -(\eta_{2I, 2J} - \eta_{2I, 2K})E_{e^{J} - e^{K}} \label{bxx2}\\ 
	[H_{I},E_{-(e^{J} + e^{K})}] &\equiv [H_{I},F_{e^{J} + e^{K}}] = (\eta_{2I, 2J} + \eta_{2I, 2K})F_{e^{J} + e^{K}} \label{bxx3}\\ 
	[H_{I},E_{-(e^{J} - e^{K})}] &\equiv [H_{I},F_{e^{J} - e^{K}}] = (\eta_{2I, 2J} - \eta_{2I, 2K})F_{e^{J} - e^{K}} \label{bxx4}\\ 
	[E_{e^{J}+e^{K}},F_{e^{J} + e^{K}}] &= -(\eta_{2K, 2K}M_{2J-1,2J} + \eta_{2J, 2J}M_{2K-1,2K}) = -(\eta_{2K 2K}H_{J} + \eta_{2J 2J}H_{K}) \label{axx}\\
	[E_{e^{J} - e^{K}},F_{e^{J}-e^{K}}] &= -(\eta_{2K, 2K}M_{2J-1,2J} - \eta_{2J, 2J}M_{2K-1,2K}) = -(\eta_{2K 2K}H_{J} - \eta_{2J 2J}H_{K}) \label{axx2}
\end{align}
We also note that the remaining commutation relations can be summarized as 
\begin{align}
[E_{\xi e^{J} + \xi' e^{K} }, E_{\delta e^{M} + \delta' e^{N} } ] &= - i \eta_{2J, 2M}(1-\xi \delta)E_{\xi'e^{K} + \delta' e^{N} } - i \eta_{2 K, 2 N}(1- \xi' \delta')E_{\xi e^{J} + \delta e^{M} }\nonumber\\ &+ i \eta_{2J, 2N}(1-\xi \delta')E_{\xi' e^{K} + \delta e^{M} } + i \eta_{2K,2N}(1- \xi' \delta)E_{\xi e^{J} + \delta' e^{N} } \label{fullCommutator}
\end{align}
Note that the coefficients of the roots $\xi, \xi', \delta, \delta'$ can take values $\pm 1$. Moreove, the lowering operator $F_{\alpha}$ is given as $F_{\alpha} = E_{-\alpha}$, where $\alpha$ is some root.
In writing these relations, we have used the fact that 
\bea
\eta_{2I, 2J}=\eta_{2I-1,2J-1}. \label{simplE}
\eea
Incidentally, \eqref{fullCommutator} can be viewed to implicitly contain \eqref{axx3}-\eqref{axx2} as well, via a slight notational re-interpretation. This is accomplished when $J=M$ and $K=N$, with the understanding that $E_{e^{J} - e^{J}}$ is to be replaced with $-i H_J$.
In any event, we now have a Cartan-Weyl-like form associated to the $SO(2n,2m)$ algebra that can be transparently used for discussing highest weight states. 

One further comment worth making here is that in the $SO(2,2)$ case, the algebra presented in \eqref{axx3}-\eqref{axx2} is the complete algebra. The rest of the commutators in \eqref{fullCommutator} become trivial.  


\subsection{Index Summary}

We have had to introduce a few different kinds of indices in the discussions so far. To avoid confusion and for quick reference, let us summarize our notation here. 

Recall that we are working with $SO(2n,2m)$. The uppercase Latin indices $A$ and $B$ go over the full range of coordinates of the Klein space, $1, \cdots , 2n+2m$. These show up in \eqref{comm}. These are further subdivided via the lowercase un-tilded Latin indices $a, b$ (which span the coordinates denoted by $x$, with signature $-1$) and the lowercase tilded Latin indices $\t{a},\t{b}$ (which span the coordinates denoted by $y$, with signature $+1$). One can relate these indices to the uppercase Latin indices as follows
\begin{align}
	A &= a \; \forall \; A \in {1,\dots,2n}\\
	A &= 2n + \t{a} \; \forall \; A \in {2n+1,\dots,2n+2m}
\end{align}

Now, we turn to the upper case Latin indices $I, J$ and $K$ of the previous subsection. These go over half the range of coordinates, ie., $I, J, K \in 1,\dots,n+m$. The full range of Klein space coordinates is then covered by $2J$ and $2J-1$ together (and similarly for $K$). These are further subdivided via the lowercase un-tilded Latin indices $j, k$ (which span \emph{half} of the coordinates $x$ and therefore label the coordinates $z, \bar{z}, r$ and $\theta$ from section \ref{sec2.1}) and the lowercase tilded Latin indices $\t{j},\t{k}$ (which span \emph{half} of the coordinates $y$, and label $w, \bar{w}, t$ and $\phi$ from section \ref{sec2.1}). The full range of the $x$ coordinates is spanned by $2j$ and $2j-1$ together (and similarly for $k$). Likewise, $y$ is spanned by $2\t{j}$ and $2\t{j}-1$ and by $2\t{k}$ and $2\t{k}-1$. One can relate the uppercase and lowercase Latin indices in this case as follows
\begin{align}
	J &= j \; \forall \; J \in {1,\dots,n}\\
	J &= n + \t{j} \; \forall \; J \in {n+1,\dots,n+m}
\end{align}
The upper case $J,K$ indices and their translation to the lower case $j,k$ and $\t{j}, \t{k}$ indices will be repeatedly used in what follows.

Finally, for completeness, we make a small comment on the Latin indices $\mu,\nu$, used in \eqref{mab}. These are simply the matrix indices on the matrix-form of the generators $J_{A B}$ of $SO(2n,2m)$. The matrices are $2(n+m) \times 2(n+m)$ dimensional in this representation, so the indices $\mu, \nu \in 1,\dots,2n+2m$. We will not need these indices again.

\subsection{Quadratic Casimir and the ``Wave'' Equation}

The quadratic Casimir for $SO(p,q)$ group can be taken as (see Appendix D for a discussion):
\begin{equation}
	c_{2} = M_{A B}M^{A B} \label{331a}
\end{equation}
where $M_{A B}$ is defined as \eqref{mab}, and the raising/lowering of indices is executed via the Klein space metric tensor $\eta_{AB}$.
In terms of the generators of $SO(p,q)$ in the Cartan-Weyl-like basis we introduced, one can express the quadratic Casimir as follows 
\begin{equation}
	\begin{split}
		c_{2} = \sum_{I = 1}^{\frac{p+q}{2}} H_{I}H^{I} + \sum_{J, K > J} E_{e^{J} + e^{K}}F^{e^{J} + e^{K}} + F_{e^{J} + e^{K}}E^{e^{J} + e^{K}}\\
		+ \sum_{J, K > J} E_{e^{J} - e^{K}}F^{e^{J} - e^{K}} + F_{e^{J} - e^{K}}E^{e^{J} + e^{K}}\label{332}
	\end{split}
\end{equation}
Here, we use the $\eta^{AB}$ to raise indices on the right hand sides of the definitions \eqref{Klein-Cartan},\eqref{KleinRaising1}, \eqref{KleinRaising2}, \eqref{KleinLowering1}, \eqref{KleinLowering2} to define the upper index objects $H^I, E^{\alpha_{J,K}}$ and $E^{-\alpha_{J,K}}$. For example,
\begin{align}	
	H^{I} &= M^{2I-1,2I}\\
	E^{e^{J} \pm e^{K}} &= \frac{1}{2}\{M^{2J-1,2K-1} + i M^{2J, 2K-1} \pm i(M^{2J-1,2K} + i M^{2J,2K})\}\label{311}\\
	F^{e^{J} \pm e^{K}} &= \frac{1}{2}\{M^{2J-1,2K-1} - i M^{2J, 2K-1} \mp i(M^{2J-1,2K} - i M^{2J,2K})\}\label{312}
\end{align}
It can be seen by using these in \eqref{332}, that we get \eqref{331a}.

The above expressions are useful because using them, we can write the Casimir in terms of the differential operators from the previous sections that capture the Klein space isometries. This Casimir plays an important role in the Klein space wave equation.  To illustrate this, we work for the rest of this and the following section, with the $\mathbb{K}^{p,q-}$ foliation from section 2. The $\mathbb{K}^{p,q+}$ case follows in a similar fashion.

The wave equation in the $AdS$-foliating coordinates is as follows (using \eqref{2.10}-\eqref{2.11}):
\begin{equation}\label{box}
	\Box\Phi = \partial^{2}_{\tau}\Phi + \frac{p+q-1}{\tau}\partial_{\tau}\Phi - \frac{1}{\tau^{2}}\nabla_{p+q-1}\Phi
\end{equation}
where we have
\begin{equation}
	\nabla_{p+q-1}\Phi = 
	\Big( \partial^{2}_{\rho}\Phi + \Big((p-1){\tanh\rho} + (q-1){\coth\rho}\Big)\partial_{\rho}\Phi - \frac{1}{\cosh^{2}\rho}\nabla^{2}_{\Omega,\; p-1}\Phi + \frac{1}{\sinh^{2}\rho}\nabla^{2}_{\tilde{\Omega}\;q-1}\Phi \Big)
\end{equation}
This allows us to separate the solution $\Phi = \Phi_{1}(\tau)\Phi_{2}(\rho,\Omega,\tilde{\Omega})$, which gives us the following set of differential equations
\begin{align}
	\frac{1}{\tau^{p+q-3}}\partial_{\tau}\tau^{p+q-1}\partial_{\tau}\Phi_{1} &= K \Phi_{1} \label{km}\\
	\nabla_{p+q-1}\Phi_{2} &= K\Phi_{2}\label{K}
\end{align}
A key point is that Casimir $c_{2}$ is the same as $\nabla_{p+q-1}$. This can be explicitly checked, and this fact will be crucial as we proceed.

\subsection{Action of Casimir on Highest Weight States}

Let us consider states that are annihilated by the raising operators (denoted by $E_{\alpha_{J,K}}$ or $E_{e^{J}\pm e^{K}}$). We will denote them by $\Phi_{++}$; these are the \emph{highest weight states}. The discussion for lowest weight and mixed weight states will follow similarly. \par
To turn \eqref{332} into a useful form, it is convenient to first write 
\begin{equation}
	E^{e^{J} \pm e^{K}} = 
	\eta^{2J, 2J}\eta^{2K, 2K}E_{e^{J} \pm e^{K}}
\end{equation}
where there is no summation in $I$ or $J$. This is a simple consequence of the fact that raising and lowering of the full $M_{A B}$ is carried out by the Klein space metric $\eta^{A B}$, which has the property \eqref{simplE}. Let us illustrate this  for $E_{e^{J}\pm e^{K}}$. This will serve as a representative for the standard manipulations in these subsections. We note that
\begin{equation}
	\begin{split}
		E^{e^{J} \pm e^{K}}  &\equiv \frac{1}{2}\{M^{2J-1,2K-1} + i M^{2J, 2K-1} \pm i(M^{2J-1,2K} + i M^{2J,2K})\}\\
		&= \frac{1}{2}\{ \eta^{2J-1, A}\eta^{2K-1, B}M_{A, B} + i \eta^{2J, A}\eta^{2K-1, B}M_{A, B} \pm i(\eta^{2J-1, A}\eta^{2K, B}M_{A,B} + i \eta^{2J, A}\eta^{2K, B}M_{A,B})\}
	\end{split}
\end{equation}
Because $\eta^{ A B}$ is diagonal, we can write this as
\begin{equation}
	\begin{split}
		E^{e^{J} \pm e^{K}} &= \frac{1}{2}\{ \eta^{2J-1, 2J-1}\eta^{2K-1, 2K-1}M_{2J-1, 2K-1} + i \eta^{2J,2J}\eta^{2K-1, 2K-1}M_{2J, 2K-1} \\ &\pm i(\eta^{2J-1, 2J-1}\eta^{2K, 2K}M_{2J-1,2K} + i \eta^{2J, 2J}\eta^{2K, 2K}M_{2J,2K})\}\label{raise}
	\end{split}
\end{equation}
From the structure of $\eta^{A B}$, given as $(-\mathbf{1}_{2n}, \mathbf{1}_{2m}) $, and recalling that $J,K \in 1,\dots,n+m$, one can note that $\eta^{2J-1, 2J-1}  = 
\eta^{2J, 2J}$, and similarly for $K$. From this, what we wanted to show immediately follows:
\begin{equation}
	E^{e^{J} \pm e^{K}} = \frac{1}{2}\eta^{2J, 2J}\eta^{2K, 2K}\{ M_{2J-1, 2K-1} + i M_{2J, 2K-1} \pm i (M_{2J-1,2K} + i M_{2J,2K})\} = \eta^{2J,2 J}\eta^{2K,2K}E_{\alpha_{J, K}}
\end{equation}

Using this gives us the following result for the Casimir: 
\begin{equation}
	\begin{split}
		c_{2} &= \sum_{I} H_{I}H_{I} + \sum_{J, K>J} \eta^{2J, 2J}\eta^{2K, 2K}(F_{e^{J} + e^{K}}E_{e^{J} + e^{K}} + E_{e^{J} + e^{K}}F_{e^{J} + e^{K}})\\ &+ \sum_{J, K>J}\eta^{2J, 2J}\eta^{2K, 2K}(F_{e^{J}- e^{K}}E_{e^{J}- e^{K}} + E_{e^{J}- e^{K}}F_{e^{J}- e^{K}})
	\end{split}
\end{equation}
Utilising commutation relations like 
\eqref{axx} we can now place the raising operators to the right (and thus directly act on the field $\Phi_{++}$)
\begin{equation}
	\begin{split}
		c_{2} &= \sum_{I} H_{I}H_{I} + \sum_{J, K>J} \eta^{2J, 2J}\eta^{2K, 2K}(2F_{e^{J} + e^{K}}E_{e^{J} + e^{K}} + [E_{e^{J} + e^{K}},F_{e^{J} + e^{K}}])\\ &+ \sum_{J, K>J}\eta^{2J, 2J}\eta^{2K, 2K}(2F_{e^{J} - e^{K}}E_{e^{J} - e^{K}} + [E_{e^{J} - e^{K}},F_{e^{J} - e^{K}}])
	\end{split}
\end{equation}
When $c_{2}$  acts on $\Phi_{++}$ the $E_{e^{J} \pm e^{K}}$ terms annihilate it. 
We can again use commutation relations like \eqref{axx} to write down $c_{2}\Phi_{++}$ only in terms of the Cartan generators $H_{I}$ as 
\begin{align}
	\begin{split}
		c_{2}\Phi_{++} &= \sum_{J}H_{J}^{2}\Phi_{++} - \sum_{J, K>J}\eta^{2J, 2J}\eta^{2K, 2K}(\eta_{2K, 2K}H_{J} + \eta_{2J, 2J}H_{K})\Phi_{++}\\ &- \sum_{J, K>J}\eta^{2J, 2J}\eta^{2K ,2K}(\eta_{2K 2K}H_{J} - \eta_{2J, 2J}H_{K})\Phi_{++}\nonumber
	\end{split}\\
	&= \sum_{J}H_{J}^{2}\Phi_{++}  - 2\sum_{J,\; K>J}\eta^{2J, 2J}H_{J}\Phi_{++}\nonumber\\
	&= \sum_{J=1}^{n+m}H_{J}^{2}\Phi_{++} + 2\sum_{J=1,\; K>J}^{n,\; n+m}H_{J}\Phi_{++} - 2\sum_{J=n+1,\; K>J}^{n+m,\; n+m}H_{J}\Phi_{++}\nonumber\\
	\implies c_{2}\Phi_{++} &= \sum_{J=1}^{n+m}\Big( H_{J}^{2} -2\eta^{2J, 2J}(n+m-J)H_{J} \Big)\Phi_{++}\label{casimir}
\end{align}
This is the action of the quadratic Casimir of $SO(2n,2m)$ on the heighest weight state. \par 
Similary, one can see that the action of $c_{2}$ on the lowest weight state $\Phi_{--}$ (annihilated by the lowering operators) is 
\begin{equation}
	c_{2}\Phi_{--} = \sum_{J=1}^{n+m}\Big(H_{J}^{2} + 2\eta^{2J, 2J}(n+m-J)H_{I}\Big)\Phi_{--}\label{K2}
\end{equation}
In fact, it is easy to see that there exists a huge class of such states one can define by considering states annihilated by various combinations of raising and lowering operators. In fact, there exists $2^r$ such choices, where $r=(n+m)(n+m-1)$ is the number of positive roots of $SO(2n,2m)$ -- the idea being that we need to impose $r$ annihilation conditions, and we can make the choice of either the positive root itself, or the corresponding negative root\footnote{Note that this reduces to the $4$ kinds of states found in \cite{Strominger} when we work with $n=m=1$.}. Of these, the highest weight and lowest weight states constitute just two.  We will present the action of $c_{2}$ on two examples of such ``mixed'' weight states and leave the rest as an exercise for the reader. The first kind $\Phi_{+-}$ (annihilated by raising operators corresponding to roots of the form $e^{J} + e^{K}$ and by lowering operators corresponding to roots of the form  $e^{J} - e^{K}$)  yield
\begin{equation}
	c_{2}\Phi_{+-} = \sum_{J=1}^{n+m}\Big(H_{J}^{2} - 2\eta^{2J, 2J}(J-1)H_{J}\Big)\Phi_{+-}\label{K3}
\end{equation}
And finally, we write down the action of $c_{2}$ on a second kind of mixed weight state $\Phi_{-+}$ (annihilated by raising operators corresponding to roots of the form $e^{J} - e^{K}$ and by lowering operators corresponding to roots of the form  $e^{J} + e^{K}$):
\begin{equation}
	c_{2}\Phi_{-+} = \sum_{J=1}^{n+m}\Big(H_{J}^{2} + 2\eta^{2J, 2J}(J-1)H_{J}\Big)\Phi_{-+}\label{K4}
\end{equation}

\section{$L$-Primary Scalars}
\noindent
In this section, we will give a solution to the set of equations \eqref{km}-\eqref{K}. The equation \eqref{km} can be trivially solved, and we can write $\Phi_{1}(\tau)$ exactly, up to the constant $K$. We will be interested in specific solutions for $\Phi_{2}$. These solutions will be obtained by imposing additional demands on $\Phi_{2}$, related to the highest/lowest/mixed weight conditions. These are natural \cite{Strominger}, in view of the fact that $\nabla_{p+q-1}$ in \eqref{K} is the Casimir $c_{2}$ \eqref{332}. 


Separating variables in the ``wave'' equation we can easily solve 
\begin{equation}
	\frac{\tau^{2}}{\tau^{p+q-1}}\partial_{\tau}\tau^{p+q-1}\partial_{\tau}\Phi_{1} = K \Phi_{1},\label{4.16}
\end{equation}
and we find
\begin{equation}\label{4.17}
	\Phi_{1}(\tau) = \tau^{-(\frac{p+q-2}{2}) \pm \sqrt{K + \frac{(p+q-2)^{2}}{4}}}
\end{equation}


Solving for $\Phi_{2}$ requires more work. We begin by writing the expressions for the raising and lowering operators in terms of the coordinates given in \eqref{1.3}-\eqref{1.10} (see Appendix A for some related formulas and notation):
\begin{align}
	E^{x,x}_{e^{j}\pm e^{k}} &= -\frac{i}{2} e^{i(\theta_{j}\pm \theta_{k})}\{(r_{j}\partial_{r_{k}} - r_{k}\partial_{r_{j}}) \pm i\frac{r_{j}}{r_{k}}\partial_{\theta_{k}} - i\frac{r_{k}}{r_{j}}\partial_{\theta_{j}}\}\\
	F^{x,x}_{e^{j}\pm e^{k}} &= \frac{i}{2} e^{-i(\theta_{j}\pm \theta_{k})}\{(r_{j}\partial_{r_{k}} - r_{k}\partial_{r_{j}}) \mp i\frac{r_{j}}{r_{k}}\partial_{\theta_{k}} + i\frac{r_{k}}{r_{j}}\partial_{\theta_{j}}\}\\
	E^{y,y}_{e^{\t{j}}\pm e^{\tilde{k}}} &= \frac{i}{2} e^{i(\phi_{\tilde{j}}\pm \phi_{\tilde{k}})}\{(t_{\tilde{j}}\partial_{t_{\tilde{k}}} - t_{\tilde{k}}\partial_{t_{\tilde{j}}}) \pm i\frac{t_{\tilde{j}}}{t_{\tilde{k}}}\partial_{\phi_{\tilde{k}}} - i\frac{t_{\tilde{k}}}{t_{\tilde{j}}}\partial_{\phi_{\tilde{j}}}\}\\
	F^{y,y}_{e^{\tilde{j}}\pm e^{\tilde{k}}} &= -\frac{i}{2} e^{-i(\phi_{\tilde{j}}\pm \phi_{\tilde{k}})}\{(t_{\tilde{j}}\partial_{t_{\tilde{k}}} - t_{\tilde{k}}\partial_{t_{\tilde{j}}}) \mp i\frac{t_{\tilde{j}}}{t_{\tilde{k}}}\partial_{\phi_{\tilde{k}}} + i\frac{t_{\tilde{k}}}{t_{\tilde{j}}}\partial_{\phi_{\tilde{j}}}\}\\
	E^{x,y}_{e^{j}\pm e^{\tilde{j}}} &= \frac{i}{2}e^{i(\theta_{j}\pm \phi_{\tilde{j}})}\{(r_{j}\partial_{t_{\tilde{j}}} + t_{\tilde{j}}\partial_{r_{j}}) \pm i \frac{r_{j}}{t_{\tilde{j}}}\partial_{\phi_{\tilde{j}}} + i\frac{t_{\tilde{j}}}{r_{j}}\partial_{\theta_{j}}\}\\
	F^{x,y}_{e^{j}\pm e^{\tilde{j}}}&= -\frac{i}{2}e^{-i(\theta_{j}\pm \phi_{\tilde{j}})}\{(r_{j}\partial_{t_{\tilde{j}}} + t_{\tilde{j}}\partial_{r_{j}}) \mp i \frac{r_{j}}{t_{\tilde{j}}}\partial_{\phi_{\tilde{j}}} - i\frac{t_{\tilde{j}}}{r_{j}}\partial_{\theta_{j}}\}
\end{align}
where $E^{x,x},\; F^{x,x}$ are the raising and lowering operators respectively for the case where the root lies among the $x$ coordinates, i.e $J = j$ and $K= k$. Similarly for $E^{y,y},\;F^{y,y}$, where the root lies among the $y$ coodinates, i.e. $J = \t{j}$ and $K = \t{k}$. The rasing and lowering operators $E^{x,y}$ and $F^{x,y}$ correspond to the roots where one of the indices $J$ and $K$ (the smaller one according to our arrangement of coordiantes \eqref{coord}) lies among the $x$ coordinates and the other lies among the $y$ coordinates. This means that $J = j$ and $K = \t{j}$. \par 

We will call states satisfying the following conditions $L$-primaries, $\Phi_{++}$\footnote{We will call states that arise by analogous lowest weight and mixed weight conditions also as $L$-primaries -- this should not cause any confusion.}.  By a slight abuse of terminology, we will also sometimes refer to them simply as highest weight states. 
\begin{align}
	H_{j}\Phi_{++} &= \lambda_{i}\Phi_{++}\\
	H_{\tilde{j}}\Phi_{++} &= \tilde{\lambda}_{\tilde{j}}\Phi_{++}\\
	E^{x,x}_{e^{j}\pm e^{k}}\Phi_{++} &= E^{y,y}_{e^{\tilde{j}}\pm e^{\tilde{k}}}\Phi_{++} = 0\\ E^{x,y}_{e^{j}\pm e^{\tilde{j}}}\Phi_{++} &= 0
\end{align}

It can be shown (Appendix B) that these equations reduce to the following set of differential equations satisfied by $\Phi_{++}$:
\begin{align}\label{4.14pan}
	\frac{1}{r_{k}}\partial_{r_k}\Phi_{++} &= \frac{1}{r_{1}}\partial_{r_1}\Phi_{++} - \frac{h}{r_{1}^{2}}\Phi_{++}\;\; \forall\;\; k \neq 1\\
	-\frac{1}{t_{\t{j}}}\partial_{t_{\tilde{j}}}\Phi_{++}&= \frac{1}{r_{1}}\partial_{r_1}\Phi_{++} - \frac{h}{r_{1}^{2}}\Phi_{++}\;\; \forall\;\; \t{j} \label{4.16pan}\\
	\frac{1}{r_{k}}\partial_{r_k}\Phi_{++} &= \frac{1}{r_{j}}\partial_{r_j}\Phi_{++}\;\; \forall \;\; k\;,\;j \neq 1\\
	\frac{1}{t_{\t{j}}}\partial_{t_{\t{j}}}\Phi_{++} &= \frac{1}{t_{\t{k}}}\partial_{t_{\t{k}}}\Phi_{++} \;\; \forall \;\; \t{j}, \t{k} \label{4.17pan}\\
	\frac{1}{r_{k}}\partial_{r_k}\Phi_{++} &= - \frac{1}{t_{\t{j}}}\partial_{t_{\t{j}}}\Phi_{++} \;\; \forall \;\; \t{j}, \;\; \text{and} \;\; k \neq 1 \label{4.108pan}
\end{align}

It follows from imposing the annihilation conditions and following the calculations in Appendix A and B, that \eqref{casimir} gives us the following value of $K$ as defined in \eqref{K},
\begin{equation}
	K = h^{2} + 2(n+m-1)h\label{K1}
\end{equation}
where we have renamed $\lambda_{1} = h$ (see Appendix B). Note that this is consistent with the usual form of the Laplacian eigenvalue on the sphere in higher dimensions. But we found it here via systematicaly working through our highest weight conditions. 

The above set of equations can be solved by parametrising these coordinates in the generalized-$AdS$-foliation in which we have written the metric in \eqref{2.10}-\eqref{2.11pan}. This is done in Appendix C, and up to an orverall normalization, the solution we obtain is:
\begin{equation}
	\Phi_{++} = \Big(\frac{r_{1}}{\tau}\Big)^{h}e^{i h \theta_{1}}\label{soln01}
\end{equation}
Single-valuedness forces $h$ to be integral. Our choice of coordinates and generators are adapted to make this expression simple.

For the lowest and mixed weight states, the structure of the differential equations we need to solve is similar: The weight conditions force all except one of the Cartan eigenvalues to be zero, and this weight fixes $K$ via the Casimir equations we presented in Section 3.5. The results can be summarized as --

\begin{itemize}
	\item For the lowest weight state $\Phi_{--}$
	\begin{align}
		K &= \lambda_{1}^{2} - 2(n+m-1)\lambda_{1} \label{Ko2}\\
		\Phi_{--} &= \Big(\frac{r_{1}}{\tau}\Big)^{\lambda_1}e^{-i \lambda_1 \theta_{1}}\label{soln02}
	\end{align}
	\item For the mixed weight state of the first kind $\Phi_{+-}$
	\begin{align}
		K &= \tilde{\lambda}_{m}^{2} - 2(n+m-1)\tilde{\lambda}_{m} \label{Ko3}\\
		\Phi_{+-} &= \Big(\frac{t_{m}}{\tau}\Big)^{\tilde \lambda_m}e^{i \tilde \lambda_m \phi_{m}}\label{soln03}
	\end{align}
	\item For the mixed weight state of the second kind $\Phi_{-+}$
	\begin{align}
		K &= \tilde{\lambda}_{m}^{2} + 2(n+m-1)\tilde{\lambda}_{m} \label{Ko4}\\
		\Phi_{-+} &= \Big(\frac{t_{m}}{\tau}\Big)^{\tilde \lambda_m}e^{-i \tilde \lambda_m \phi_{m}}\label{soln04}
	\end{align}
\end{itemize}
The solutions \eqref{soln01},\eqref{soln02},\eqref{soln03} and \eqref{soln04} were obtained by solving the appropriate differential equations given in Appendix B. The key simplification as we noted is that all except one of the Cartans vanish. This Cartan lies among the $x$-coordinates for the first two and among the $y$-soordinates for the last two. The method for solving these equations is sketched out in Appendix C. Note that the $r_m/\tau$ and $t_m/\tau$ factors can be made completely explicit, if we choose coordinates on the underlying foliation (these coordinates are presented in Appendix C).




By acting with the ladder operators on the primaries defined above, we can straightforwardly construct descendant states as well. These span the full representation. We will write down examples in the first descendant level, starting with the highest weight state \eqref{soln01}. The descendant states are constructed by the action of the lowering operators (as defined in \eqref{KleinLowering1}-\eqref{KleinLowering2} ) on $\Phi_{++}$. \par 
One can explicitly check (from the expressions in \eqref{3.14},\eqref{3.16} and \eqref{3.18} ), that the only non-vanishing descendant states of the first level are given by $F^{x,x}_{e^{1} \pm e^{k}}\Phi_{++}\; \forall k \in 2,\dots,n$ and $F^{x,y}_{e^{1} \pm e^{\t{j} }}\Phi_{++}\; \forall \t{j} \in 1,\dots,m$. Their explicit expressions take the form
\begin{align}
	F^{x,x}_{e^{1} \pm e^{k}}\Phi_{++} = - i h \frac{r_{k}}{r_{1}}e^{-i(\theta_{1} \pm \theta_{k}) }\Phi_{++} \label{desc1} \\
	F^{x,y}_{e^{1} \pm e^{\t{j} }}\Phi_{++} = - i h \frac{t_{\t{j} }}{r_{1}}e^{-i(\theta_{1} \pm \phi_{\t{j} }) }\Phi_{++} \label{desc2}
\end{align}
Higher descendants can also be constructed. 
Similarly one can construct (higher) descendant levels for $\Phi_{++}$, and likewise for $\Phi_{--}$, $\Phi_{+-}$ and $\Phi_{-+}$. Note that if one starts with $\Phi_{+-}$ \eqref{soln03}, then the nonvanishing first-level descendant states would be given by the action of \eqref{3.16} and \eqref{3.18}, while \eqref{3.14} would annihilate $\Phi_{+-}$. 

\section{$H$-Primary Scalars}

In this section, we construct states which are annihilated by the raising operators in a basis in which some of the Cartan generators are boost-type. This will correspond to particles that emerge at some point on the celestial $S^{p-1}\times S^{q-1}$. We will make the choice of the generators adapted to the location of this point.

We will choose the point where the particle exits to be the in the  $x_{1}$-$x_{2}$ plane of the ``minus'' signature coordinates and the $y_{1}$-$y_{2}$ plane of the ``plus'' signature coordinates. This can always be done by choosing our coordinates appropriately without loss of generality. The $x$-coordinate location where the particle emerges can therefore be captured by an angle in the $x_1$-$x_2$ plane:
\begin{equation}
	R_{x}(\{\hat{\theta_{i}}\}) = 
	\begin{pmatrix}
		\cos\hat{\theta}_{1} & \sin\hat{\theta}_{1} & 0 & 0 & \dots & 0 & 0\\
		-\sin\hat{\theta}_{1} & \cos\hat{\theta}_{1} & 0 & 0 & \dots & 0 & 0\\
		0 & 0 & 1 & 0 & \dots & 0 & 0\\
		0 & 0 & 0 &1 & \dots & 0 & 0\\
		\vdots & \vdots & \vdots & \vdots & \vdots & \vdots & \vdots\\
		0 & 0 & 0 & \dots & 0 & 1 & 0\\
		0 & 0 & 0 & \dots & 0 & 0 & 1
	\end{pmatrix}\label{matrix2}
\end{equation}
Similarly, we will have a rotation matrix for the $y$ coordinates, with $\hat{\theta}_{1}$ replaced by $\hat{\phi}_{1}$. Once the rotations are done, the $x_1, x_2, y_1, y_2$ coordinates should really be viewed as primed coordinates. But while constructing generators in the next sub-section, we will suppress the primes. The use of the rotations is simply to clarify the angles of the momenta as we will see in subsection \ref{momsub} - if we do not include these rotations, the particle will emerge {\em along} the $x_1$ direction and $y_1$ direction, which makes the notation a bit too slim to be transparent. But the key physics is happening in the choice of Cartan generators that mix the signatures that we will present in subsection \ref{boostcartan}.

As we will see when we discuss the plane wave solutions in an upcoming subsection, the Cartans corresponding to the generators that mix the $x_{1},x_{2}$ and $y_{1},y_{2}$ planes are the only ones that play an important role in the discussion. In the present section, for the most part, we will treat the rest of the Cartan generators to be rotation generators in the two signatures separately. Our main observations remain unchanged even if we consider signature-mixing among the remaining generators. In a later subsection, we will also discuss such an alternate choice for the remaining Cartan generators. 

\subsection{Explicit $H$-Generators}\label{boostcartan}


To write down the Cartan generators in these new coordinates, we split the coordinates in the following arrangement\footnote{As mentioned, we are suppressing the primes on the four rotated coordinates.}
\begin{equation}
	X = \{ x_{1},y_{1},x_{2}, y_{2}, x_{3},\dots, x_{2n}, y_{3},\dots y_{2m}\}
\end{equation}
We will have $6$ different types of ladder operators, depending on which group the coordinates belong to -- both coordinates belonging to $x_{1},\dots,y_{2}$ (\eqref{raise1min} and \eqref{lower1min}), one from $x_{1},\dots,y_{2}$ and one from $x_{3},\dots,x_{2n}$ (\eqref{raise3min} and \eqref{lower3min}), one from $x_{1},\dots,y_{2}$ and one from $y_{3},\dots,y_{2m}$ (\eqref{raise2min} and \eqref{lower2min}), both from $x_{3},\dots,x_{2n}$ (\eqref{raise4min} and \eqref{lower4min}), both from $y_{3},\dots,y_{2m}$ (\eqref{raise5min} and \eqref{lower5min}, and finally, one from $x_{3},\dots,x_{2n}$ and one from $y_{3},\dots,y_{2m}$ (\eqref{raise6min} and \eqref{lower6min}). The Cartans are picked by choosing coordinates, either in $x_{1},\dots,y_{2}$ (\eqref{boostCart1} and \eqref{boostCart2}), or $x_{3},\dots,x_{2n}$ \eqref{rotCart1}, or in $y_{3},\dots,y_{2m}$ \eqref{rotCart2}. We emphasize that the ugliness in our listing below is purely notational -- the idea is simple and as we outlined in the introduction to this section. We will show a more presentable set of generators in a later subsection by doing some (conceptually unnecessary) rotations/boosts.

In any event, we list the generators below for completeness.  The new Cartan generators are
\begin{align}
	\t{H}_{1} &= i M_{1 \t{1}} \label{boostCart1}\\
	\t{H}_{2} &= i M_{2 \t{2}} \label{boostCart2}\\
	\t{H}_{j} &= M_{2j-1, 2j} \;\; \forall \; j \in 2,\dots,n \label{rotCart1}\\
	\t{H}_{\t{j} } &= - M_{2\t{j} -1,2\t{j}} \;\; \forall \; \t{j} \in 2,\dots,m \label{rotCart2}
\end{align}
The key point here is that unlike before, now two of our Cartans are boost-like. The raising operators are 
\begin{align}
	\t{E}^{(1)}_{e^{J}\pm e^{K}} &=  \frac{1}{2}\{ (M_{J K} \pm M_{\t{J},\t{K}}) + (M_{K \t{J}} \mp M_{J \t{K} }) \} \;\; ; J = 1\;\&\; K = 2 \label{raise1min}\\ 
	\t{E}^{(2)}_{e^{J} \pm e^{\t{ j } } }&= -\frac{i}{2}\{ (M_{J, 2\t{j}-1} - M_{\t{J}, 2\t{j}-1 } ) \mp i(M_{J, 2\t{j} } - M_{\t{J}, 2\t{j} } )\} \forall J \in 1,2 \; \& \;  \t{j} \in 2,\dots,m \label{raise2min} \\
	\t{E}^{(3)}_{e^{J} \pm e^{j}} &= \frac{1}{2}\{ (M_{J, 2j-1} + M_{2j-1,\t{J} }) \pm i(M_{J, 2j} + M_{2j,\t{J} } )  \} \forall J \in 1,2 \; \& \;  \t{j} \in 2,\dots,n \label{raise3min} \\
	\t{E}^{(4)}_{e^{j} \pm e^{k}} &= \frac{1}{2}\{ (M_{2j-1,2k-1} + i M_{2j,2k-1}) \pm i(M_{2j-1,2k} + i M_{2j,2k} )\}\forall k > j \in 2,\dots,n \label{raise4min} \\
	\t{E}^{(5)}_{e^{\t{j}} \pm e^{\t{k}} } &= -\frac{1}{2}\{(M_{2\t{j} -1, 2\t{k} -1 }  - i M_{2\t{j} , 2\t{k} -1 } ) \mp i(M_{2\t{j} -1, 2\t{k}}  - i M_{2\t{j} , 2\t{k}} ) \} \forall \t{k} > \t{j} \in 2,\dots,m \label{raise5min}\\
	\t{E}^{(6)}_{e^{j} \pm e^{\t{j}}} &= \frac{1}{2}\{(M_{2j-1,2\t{j} - 1 } + i M_{2j, 2\t{j} -1}) \mp i (M_{2j-1,2\t{j} } + i M_{2j,2\t{j} }) \}\forall j \in 2,\dots,n \;\&\; \t{j} \in 2,\dots,m \label{raise6min}
\end{align}
The corresponding lowering opeators are
\begin{align}
	\t{F}^{(1)}_{e^{J}\pm e^{K}} &=   \frac{1}{2}\{ (M_{J K} \pm M_{\t{J},\t{K}}) - (M_{K \t{J}} \mp M_{J \t{K} }) \} ; J =1 \;\&\; K=2 \label{lower1min} \\ 
	\t{F}^{(2)}_{e^{J} \pm e^{\t{ j } } }&=  \frac{i}{2}\{ (M_{J, 2\t{j}-1} + M_{\t{J}, 2\t{j}-1 } ) \pm i(M_{J, 2\t{j} } + M_{\t{J}, 2\t{j} } )\} \forall J \in 1,2 \; \& \;  \t{j} \in 2,\dots,m \label{lower2min} \\
	\t{F}^{(3)}_{e^{J} \pm e^{j}} &= \frac{1}{2}\{ (M_{J, 2j-1} - M_{2j-1,\t{J} }) \mp i(M_{J, 2j} - M_{2j,\t{J} } )  \} \forall J \in 1,2 \; \& \;  \t{j} \in 2,\dots,n \label{lower3min} \\
	\t{F}^{(4)}_{e^{j} \pm e^{k}} &= \frac{1}{2}\{ (M_{2j-1,2k-1} - i M_{2j,2k-1}) \mp i(M_{2j-1,2k} - i M_{2j,2k} )\}\forall k > j \in 2,\dots,n \label{lower4min} \\
	\t{F}^{(5)}_{e^{\t{j}} \pm e^{\t{k}} } &= \frac{1}{2}\{(M_{2\t{j} -1, 2\t{k} -1 }  + i M_{2\t{j} , 2\t{k} -1 } ) \pm i(M_{2\t{j} -1, 2\t{k}}  + i M_{2\t{j} , 2\t{k}} ) \} \forall \t{k} > \t{j} \in 2,\dots,m \label{lower5min}\\
	\t{F}^{(6)}_{e^{j} \pm e^{\t{j}}} &= \frac{1}{2}\{(M_{2j-1,2\t{j} - 1 } - i M_{2j, 2\t{j} -1}) \pm i (M_{2j-1,2\t{j} } - i M_{2j,2\t{j} }) \}\forall j \in 2,\dots,n \;\&\; \t{j} \in 2,\dots,m \label{lower6min}
\end{align}
As a simple check, one can count the total number of generators in \eqref{boostCart1}-\eqref{lower6min}. It turns out to be $(m+n)(2(m+n)-1)$, as required.

We note the commutation relations satisfied by these operators below.\newline
For $\t{E}^{(1)}$ and $\t{F}^{(1)}$
\begin{align}
	[\t{H}_{I}, \t{E}^{(1)}_{e^{J} \pm e^{K}}] &= -(\eta_{I J} \pm \eta_{I K}) \t{E}^{(1)}_{e^{J} \pm e^{K}}\label{commute1min}\\
	[\t{H}_{I}, \t{F}^{(1)}_{e^{J} \pm e^{K}}] &= (\eta_{I J} \pm \eta_{I K}) \t{F}^{(1)}_{e^{J} \pm e^{K}}\label{commute2min}\\
	[\t{E}^{(1)}_{e^{J} \pm e^{K}}, \t{F}^{(1)}_{e^{J} \pm e^{K}}] &= -(\eta_{K K} \t{H}_{J} \pm \eta_{J J}\t{H}_{K})\label{commute3min}
\end{align}
For $\t{E}^{(2)}$ and $\t{F}^{(2)}$
\begin{align}
	[\t{H}_{I}, \t{E}^{(2)}_{e^{J} \pm e^{\t{j}}}] &= -\eta_{I J}\t{E}^{(2)}_{e^{J} \pm e^{\t{j}}}\label{commute4min}\\
	[\t{H}_{\t{k}}, \t{E}^{(2)}_{e^{J} \pm e^{\t{j}}}] &= \mp \eta_{2\t{j}, 2\t{k}}\t{E}^{(2)}_{e^{J} \pm e^{\t{j}}}\label{commute5min}\\
	[\t{H}_{I}, \t{F}^{(2)}_{e^{J} \pm e^{\t{j}}}] &= \eta_{I J}\t{F}^{(2)}_{e^{J} \pm e^{\t{j}}}\label{commute6min}\\
	[\t{H}_{\t{k}}, \t{F}^{(2)}_{e^{J} \pm e^{\t{j}}}] &= \pm \eta_{2\t{j}, 2\t{k}}\t{F}^{(2)}_{e^{J} \pm e^{\t{j}}}\label{commute7min}\\
	[\t{E}^{(2)}_{e^{J} \pm e^{\t{j} }}, \t{F}^{(2)}_{e^{J} \pm e^{\t{j} }}] &= -(\eta_{2\t{j}, 2\t{j}}\t{H}_{J} \pm \eta_{J,J}\t{H}_{\t{j}})\label{commute8min}
\end{align}
For $\t{E}^{(3)}$ and $\t{F}^{(3)}$
\begin{align}
	[\t{H}_{I}, \t{E}^{(3)}_{e^{J} \pm e^{j}}] &= -\eta_{I J}\t{E}^{(3)}_{e^{J} \pm e^{j}}\label{commute9min}\\
	[\t{H}_{\t{k}}, \t{E}^{(3)}_{e^{J} \pm e^{j}}] &= \mp \eta_{2 j, 2 k}\t{E}^{(3)}_{e^{J} \pm e^{j}}\label{commute10min}\\
	[\t{H}_{I}, \t{F}^{(3)}_{e^{J} \pm e^{j}}] &= \eta_{I J}\t{F}^{(3)}_{e^{J} \pm e^{j}}\label{commute11min}\\
	[\t{H}_{\t{k}}, \t{F}^{(3)}_{e^{J} \pm e^{j}}] &= \pm \eta_{2j, 2k}\t{F}^{(3)}_{e^{J} \pm e^{j}}\label{commute12min}\\
	[\t{E}^{(3)}_{e^{J} \pm e^{j }}, \t{F}^{(3)}_{e^{J} \pm e^{j }}] &= -(\eta_{2j, 2j}\t{H}_{J} \pm \eta_{J,J}\t{H}_{j})\label{commute13min}
\end{align}
For $\t{E}^{(4)}$ and $\t{F}^{(4)}$
\begin{align}
	[\t{H}_{i}, \t{E}^{(4)}_{e^{\t{j}} \pm e^{\t{k}}}] &= -(\eta_{2i,2j} \pm \eta_{2i,2k})\t{E}^{(4)}_{e^{j} \pm e^{k}} \label{commute14min}\\
	[\t{H}_{i}, \t{F}^{(4)}_{e^{j} \pm e^{k}}] &= (\eta_{2i,2j} \pm \eta_{2i,2k})\t{F}^{(4)}_{e^{j} \pm e^{k}} \label{commute15min}\\
	[\t{E}^{(4)}_{e^{j} \pm e^{k}}, \t{F}^{(4)}_{e^{j} \pm e^{k}}] &= -(\eta_{2k, 2 k}\t{H}_{j} \pm \eta_{2 j, 2j}\t{H}_{k}) \label{commute16min}
\end{align}
For $\t{E}^{(5)}$ and $\t{F}^{(5)}$
\begin{align}
	[\t{H}_{\t{i}}, \t{E}^{(5)}_{e^{\t{j}} \pm e^{\t{k}}}] &= -(\eta_{2\t{i},2\t{j}} \pm \eta_{2\t{i},2\t{k}})\t{E}^{(5)}_{e^{\t{j}} \pm e^{\t{k}}} \label{commute17min}\\
	[\t{H}_{\t{i}}, \t{F}^{(5)}_{e^{\t{j}} \pm e^{\t{k}}}] &= (\eta_{2\t{i},2\t{j}} \pm \eta_{2\t{i},2\t{k}})\t{F}^{(5)}_{e^{\t{j}} \pm e^{\t{k}}} \label{commute18min}\\
	[\t{E}^{(5)}_{e^{\t{j}} \pm e^{\t{k}}}, \t{F}^{(5)}_{e^{\t{j}} \pm e^{\t{k}}}] &= -(\eta_{2\t{k}, 2\t{k}}\t{H}_{\t{j}} \pm \eta_{2\t{j}, 2\t{j}}\t{H}_{\t{k}}) \label{commute19min}
\end{align}
For $\t{E}^{(6)}$ and $\t{F}^{(6)}$
\begin{align}
	[\t{H}_{i}, \t{E}^{(6)}_{e^{j} \pm e^{\t{j}}}] &= -\eta_{2i 2j}\t{E}^{(6)}_{e^{j} \pm e^{\t{j}}}\label{commute20min}\\
	[\t{H}_{\t{k}}, \t{E}^{(6)}_{e^{j} \pm e^{j}}] &= \mp \eta_{2 \t{k}, 2 \t{j}}\t{E}^{(6)}_{e^{j} \pm e^{\t{j}}}\label{commute21min}\\
	[\t{H}_{i}, \t{F}^{(6)}_{e^{j} \pm e^{\t{j}}}] &= \eta_{2i 2j}\t{F}^{(6)}_{e^{j} \pm e^{\t{j}}}\label{commute22min}\\
	[\t{H}_{\t{k}}, \t{F}^{(6)}_{e^{j} \pm e^{\t{j}}}] &= \pm \eta_{2\t{k}, 2\t{j}}\t{F}^{(6)}_{e^{j} \pm e^{\t{j}}}\label{commute23min}\\
	[\t{E}^{(6)}_{e^{j} \pm e^{\t{j}}}, \t{F}^{(6)}_{e^{j} \pm e^{\t{j}}}] &= -(\eta_{2\t{j}, 2\t{j}}\t{H}_{j} \pm \eta_{2j,2j}\t{H}_{\t{j}})\label{commute24min}
\end{align}
With these commutation relations at hand, we can proceed to solve the Casimir equation and obtain the highest weight state.
\subsection{Highest Weight State}
The highest weight state is defined as follows
\begin{align}
	\t{H}_{1}\Phi_{h} &= h_{1}\Phi_{h} \label{eigenval1min}\\
	\t{H}_{2}\Phi_{h} &= h_{2}\Phi_{h} \label{eigenval2min}\\
	\t{H}_{j}\Phi_{h} &= \t{h}_{j}\Phi_{h} \;\; \forall \; j \in 2,\dots,n \label{eigenval3min}\\
	\t{H}_{\t{j} }\Phi_{h} &= \t{h}_{\t{j} }\Phi_{h} \;\; \forall \; \t{j} \in 2,\dots,m \label{eigenval4min}\\
	\t{E}^{(1)}_{e^{J}\pm e^{K}}\Phi_{h} &=  0 \;\; ; J = 1\;\&\; K = 2 \label{ann1min}\\ 
	\t{E}^{(2)}_{e^{J} \pm e^{\t{ j } } }\Phi_{h}&= 0 \;\; \forall J \in 1,2 \; \& \;  \t{j} \in 2,\dots,m \label{ann2min} \\
	\t{E}^{(3)}_{e^{J} \pm e^{j}}\Phi_{h} &= 0 \;\; \forall J \in 1,2 \; \& \;  \t{j} \in 2,\dots,n \label{ann3min} \\
	\t{E}^{(4)}_{e^{j} \pm e^{k}}\Phi_{h} &= 0\;\; \forall k > j \in 2,\dots,n \label{ann4min} \\
	\t{E}^{(5)}_{e^{\t{j}} \pm e^{\t{k}} }\Phi_{h} &=0\;\; \forall \t{k} > \t{j} \in 2,\dots,m \label{ann5min}\\
	\t{E}^{(6)}_{e^{j} \pm e^{\t{j}}}\Phi_{h} &= 0\;\; \forall j \in 2,\dots,n \;\&\; \t{j} \in 2,\dots,m \label{ann6min}
\end{align}
After translating these into differential equations satisfied by $\Phi_{h}$ we find -- \\
From \eqref{ann1min}
\begin{align}
	\frac{1}{x_{2}}\frac{\partial \Phi_{h}}{\partial x_{2} } = - \frac{1}{y_{2}}\frac{\partial \Phi_{h}}{\partial y_{2} } = \frac{1}{(x_{1} - y_{1})}\Big(\frac{\partial}{\partial x_{1} } + \frac{\partial}{\partial y_{1} }\Big)\Phi_{h} \label{mincoond1}
\end{align}
From \eqref{ann2min}
\begin{align}
	-\frac{1}{y_{2\t{j}-1 } }\frac{\partial \Phi_{h}}{\partial y_{2\t{j} -1 } } = -\frac{1}{y_{2\t{j}} }\frac{\partial \Phi_{h}}{\partial y_{2\t{j}} } = \frac{1}{x_{J} - y_{\t{J}}}\Big(\frac{\partial}{\partial x_{J} } + \frac{\partial}{\partial y_{\t{J} } }\Big)\Phi_{h} \;\; \forall J = 1,2 \; \& \; \t{j} \in 2,\dots,m \label{mincoond2}
\end{align}
From \eqref{ann3min}
\begin{align}
	\frac{1}{x_{2j-1 } }\frac{\partial \Phi_{h}}{\partial x_{2j -1 } } = \frac{1}{x_{2j} }\frac{\partial \Phi_{h}}{\partial x_{2j} } = \frac{1}{x_{J} - y_{\t{J}}}\Big(\frac{\partial}{\partial x_{J} } + \frac{\partial}{\partial y_{\t{J} } }\Big)\Phi_{h} \;\; \forall J = 1,2 \; \& \; j \in 2,\dots,n \label{mincoond3}
\end{align}
From \eqref{ann4min}
\begin{align}
	\frac{1}{x_{2k-1} }\frac{\partial \Phi_{h}}{\partial x_{2k-1} } = \frac{1}{x_{2 k} }\frac{\partial \Phi_{h}}{\partial x_{2 k}} = \frac{1}{(x_{2j}- i x_{2j-1} )}\Big(\frac{\partial}{\partial x_{2 j} } - i\frac{\partial}{\partial x_{2j-1} } \Big)\Phi_{h} \;\; \forall k > j \in 2,\dots, n \label{mincoond4}
\end{align}
From \eqref{ann5min}
\begin{align}
	\frac{1}{y_{2\t{k}-1 }}\frac{\partial \Phi_{h}}{\partial y_{2\t{k}-1 }} = \frac{1}{y_{2\t{k}}}\frac{\partial \Phi_{h}}{\partial y_{2\t{k}}} = \frac{1}{(y_{2\t{j} } + i y_{2\t{j}-1}) }\Big(\frac{\partial}{\partial y_{2\t{j} } } + i \frac{\partial}{\partial y_{2\t{j}-1} } \Big)\Phi_{h} \;\; \forall \t{k} > \t{j} \in 2,\dots,m \label{mincoond5}
\end{align}
From \eqref{ann6min}
\begin{align}
	-\frac{1}{y_{2\t{j}-1 }}\frac{\partial \Phi_{h}}{\partial y_{2\t{j}-1 }} = -\frac{1}{y_{2\t{j}}}\frac{\partial \Phi_{h}}{\partial y_{2\t{j}}} = \frac{1}{(x_{2j}- i x_{2j-1} )}\Big(\frac{\partial}{\partial x_{2 j} } - i\frac{\partial}{\partial x_{2j-1} } \Big)\Phi_{h} \;\; \forall \t{j} \in 2,\dots,m \; \& j \in 2,\dots,n \label{mincoond6}
\end{align}
The expressions \eqref{mincoond1}-\eqref{mincoond6} imply that
\begin{align}
	\t{H}_{1}\Phi_{h} &= h\Phi_{h} \label{mincoond7}\\
\t{H}_{2}\Phi_{h} &= 0 \label{mincoond8}\\
\t{H}_{j}\Phi_{h} &= 0 \;\; \forall \; j \in 2,\dots,n \label{mincoond9}\\
\t{H}_{\t{j} }\Phi_{h} &= 0 \;\; \forall \; \t{j} \in 2,\dots,m \label{mincoond10}
\end{align}
Together, \eqref{mincoond1}-\eqref{mincoond10} imply that $\Phi_{h}$ is dependent only on $x_{1}, x_{2}, y_{1}, y_{2}$. The equation satisfied by $\Phi_{h}$ reduces to the following 
\begin{align}
\partial_{\bar{\theta}_{1}}\Phi_{h}&=-\frac{h r_{1} \sin \bar{\theta}_{1}}{r_{1}\cos \bar{\theta}_{1} - t_{1}\cos \bar{\phi}_{1}} \label{ultMIN1}\\
\partial_{\bar{\phi}_{1}}\Phi_{h}&=\frac{h t_{1} \sin \bar{\phi}_{1}}{r_{1}\cos \bar{\theta}_{1} - t_{1}\cos \bar{\phi}_{1}} \label{ultMIN2}
\end{align}
where we have used the notation $
\theta_{1} - \hat{\theta}_{1} \equiv \bar{\theta}_{1}, \ 
\phi_{1} - \hat{\phi}_{1} \equiv \bar{\phi}_{1}$. 
It is easy to see that the solution of \eqref{ultMIN1}-\eqref{ultMIN2} is 
\begin{equation}
\Phi_{h} = C(h)(t_{1}\cos(\hat{\phi}_{1}-\phi_{1}) - r_{1}\cos(\hat{\theta}_{1} - \theta_{1}))^{h}.\label{solutionHmin}
\end{equation}
In terms of 
\begin{equation}
z = t_{1}\cos(\hat{\phi}_{1}-\phi_{1}) - r_{1}\cos(\hat{\theta}_{1} - \theta_{1})\label{z2} 
\end{equation}
we can write $\Phi_{h}$ as a Mellin transform
\begin{equation}
\Phi_{h} \equiv \int_{0}^{\infty} \mathrm{d\omega}\; \omega^{-h-1} e^{i \omega z} = \Gamma(-h)e^{i \frac{\pi h}{2}}z^{h} \label{mellinIntmin}
\end{equation}
where we have also chosen a normalization. This representation will be useful in the next section.

\subsection{Connection with Particle Momentum}\label{momsub}

The above Mellin transform makes a connection between our primary states and ordinary plane wave states in the Klein space. Consider (as in Section 3) a coordinate system $(\{x_{a}\},\{y_{\t{a}}\})$. Let us also note that a massless particle's momentum fixes a unique point on the celestial sphere. Since it is a massless solution, we require that the norm of the momentum be $0$. Without any loss of generality, we can orient our coordinate axes in such a way that the momentum vector $\vec{p}$ lies on only $2$ out of the $n+m$ Cartan planes, with its projection on any other plane being $0$. 
Hence, we can choose the momentum vector to be
\begin{equation}
	\vec{p} = (p_{1}e^{i\hat{\theta}_{1}},p_{1}e^{-i\hat{\theta}_{1}},0, \dots, 0,q_{1}e^{ i\hat{\phi}_{1}},q_{1}e^{-i \hat{\phi}_{1}},0,\dots,0)
\end{equation}
The null constraint on this vector is as follows
\begin{equation}
	\vec{p}^{2} = \sum_{i = 1}^{n}-p_{i}^{2} + \sum_{\t{i}=1}^{m}q_{\t{i}}^{2} = 0 
\end{equation}
It forces $p_{1} = q_{1} \equiv \omega $. Hence the momentum vector is
\begin{equation}
	\vec{p} = \omega(e^{i \hat{\theta}_{1}},e^{-i \hat{\theta}_{1}},0,\dots,0,e^{ i\hat{\phi}_{1}},e^{-i \hat{\phi}_{1}},0,\dots,0)\label{pfinal2}
\end{equation}
Now, in this coordinate system, a general point in the Klein space is given by 
\begin{equation}
	\vec{X} = (r_{1}e^{i\theta_{1}},r_{1}e^{-i \theta_{1}},\dots,r_{n}e^{i \theta_{n}},r_{n}e^{-i \theta_{n}},t_{1}e^{i \phi_{i}},t_{1}e^{-i \phi_{i}},\dots,t_{n}e^{i \phi_{n}},t_{m}e^{-i \phi_{m}})
\end{equation}
We can evaluate $\vec{p}.\vec{X}$ to be
\begin{equation}
	\vec{p}.\vec{X} = -r_{1}\cos(\theta_{1} - \hat{\theta}_{1}) + t_{1}\cos(\phi_{1} - \hat{\phi}_{1})
\end{equation}
Note that this is exactly the expression $z$, eqn \eqref{z2}. So we can understand our primary states from the last section as Mellin transforms of plane wave states:
\begin{equation}
	\Phi_{h} = \int_{0}^{\infty}d\omega \omega^{-h-1}e^{i\omega \vec{p}.\vec{X}}
\end{equation}
Because $\vec{p}.\vec{X}$ is a scalar, this expression is more transparent. \par
Therefore what we have done geometrically in defining our $H$-primaties earlier in this section is to align the coordinate system suitably with the particle momentum. The Cartan-Weyl basis is chosen to reflect this. We rotated our coordinates to $\hat{\theta}$ and $\hat{\phi}$ earlier, simply to ensure that the momenta are not aligned with the axes.

\subsection{Relating $L$ and $H$ Primaries}

We can relate the $L$-primaries and their descendants to $H$-primaries and their descendants\footnote{In light of our observations in this paper, a more natural nomenclature for $L$ (and $H$) primaries is $rotation$-$like$ (and $boost$-$like$) $primaries$. But we will stick to the nomenclature introduced in \cite{Strominger} to avoid confusion.}.  In this subsection we will simply show how the primaries in both languages are related. We start by defining a transform
\begin{equation}\label{OppIntMin}
	\Phi_{\{m+, m-\}} = \int_{0}^{2\pi} \int_{0}^{2\pi}\mathrm{d\chi^{+}}\mathrm{d\chi^{-} }e^{-i m^{+} \chi^{+}- i m^{-}\chi^{-}}\Phi_{h}
\end{equation}
where $\chi^{\pm} = \hat{\theta}_{1} \pm \hat{\phi}_{1} $.
Each of angles $\hat{\theta}_{1}$ and $\hat{\phi}_{1}$ have a periodicity of $2\pi$, so $\chi^{\pm}$ have the periodicity
\begin{equation*}
	(\chi^{+},\chi^{-}) \sim (\chi^{+} \pm 2\pi,\chi^{-}\pm 2\pi) \sim (\chi^{+} \pm 2\pi,\chi^{-}\mp 2\pi)
\end{equation*}
We require our mode functions to be single valued on the celestial sphere, so
\begin{equation}
	m^{+}\pm m^{-}\in \mathbb{Z}
\end{equation}

We claim that we can choose $m^+$ and $m^-$  so that \eqref{OppIntMin} relates a highest weight $H$-primary to a highest weight $L$-primary. To demonstrate this, it is useful to first observe the action of the following generators:
\begin{align}
	(H_{1} \mp H_{\t{1}} )\Phi_{\{m+, m-\}} &= - 2 m^{\pm}\Phi_{\{m+, m-\}} \label{m1min}\\
	H_{j}\Phi_{\{m+, m-\}} = H_{\t{j}}\Phi_{\{m+, m-\}} &= 0 \;\; \forall j > 1 \; \& \; \t{j} > 1 \label{m2min}
\end{align}
These follow from direct calculation using the explicit forms of the generators (as given in sections 3 and 4) on \eqref{OppIntMin}.
\par

With this in place, we can relate the two highest weigth states. We  act with the Casimir on  \eqref{OppIntMin}. Since the $\Phi_{h}$ on the right hand side is highest weight, we have\footnote{Using \eqref{mincoond7}-\eqref{mincoond10}, the explicit form of the $H$-Casimir is presented in \eqref{casMIN}. To write the Casimir in terms of the Cartan generators and ladder operators, we need to turn to \eqref{boostCart1}-\eqref{lower6min}, and identify the correct generators with raised indices, keeping track of signs. This is done explicitly in Appendix F. But the result \eqref{nabla2min} follows from general principles.} 
\begin{equation}
	\nabla \Phi_{\{m+, m-\}} = (h^{2} + 2(n+m-1)h)\Phi_{\{m+, m-\}}\label{nabla2min}
\end{equation} 
To find that particular mode that corresponds to the $L$-primary highest weight state, we demand that (for the particular choice of $m$'s) it is annihilated by the raising operators in the Cartan basis in Section 4. We denote that particular mode by $\Phi_{\{m_{h}+, m_h-\}}$. The equation satisfied by $\Phi_{\{m_{h}+, m_h-\}}$ then shall be given by \eqref{casimir}:
\begin{equation}
	\nabla \Phi_{\{m_{h}+, m_h-\}} =  \sum_{J=1}^{n+m}\Big( H_{J}^{2} -2\eta^{2J, 2J}(n+m-J)H_{J} \Big)\Phi_{\{m_{h}+, m_h-\}}
\end{equation}
Now using \eqref{m1min}-\eqref{m2min}, we see by explicit calculation that 
\begin{eqnarray}
	\nabla \Phi_{\{m_{h}+, m_h-\}} = C \ \Phi_{\{m_{h}+, m_h-\}}, 
\end{eqnarray}
where
\begin{eqnarray}
	C= (m_h^{+} + m_h^{-})^{2} + (m_h^{+} - m_h^{-})^{2} - 2(n+m-1)(m_h^{+} + m_h^{-})- 2(m-1)(m_h^{+} - m_h^{-}). \nonumber
\end{eqnarray}
Together with the fact that the $L$-primaries depend on only the $\theta$'s or $\phi$'s but not both\footnote{We have chosen them to depend only on $\theta$'s by convention throughout this paper.}, this leads us to the following relation, by comparision with \eqref{nabla2min}:
\begin{equation}
	m_h^{+} = m_h^{-} = -\frac{h}{2}  \label{relationMIN}
\end{equation}
This corresponds to the highest weight state in the basis of Sections 3-4. 

In fact performing the integral \eqref{OppIntMin}\footnote{This can be done via a minor variation of the calculation presented in Appendix A of \cite{Strominger} explicitly, so we will skip the details.} explicitly, we can show that \eqref{soln01} emerges:
\begin{equation}\label{OppIntHighMIN}
	\begin{split}
		\Phi_{\{-\frac{h}{2}, -\frac{h}{2}\}} &= \int_{0}^{2\pi} \mathrm{d\chi^{+}}\mathrm{d\chi^{-}}e^{-i m_h^{+}\chi^{+} - i m_h^{-}\chi^{-}}\Phi_{h}\\ &= 2^{-h}4\pi^{2}e^{ i h \theta_{1}}\Gamma(-h)e^{i \frac{\Pi h}{2}}r_{1}^{h} \propto e^{i h\theta_{1}}r^{h}_{1} = \Phi_{0}
	\end{split}
\end{equation}
Similarly, one can perform the integration by making appropriate choices for $m^{\pm}$ to obtain the lowest weight and mixed weight states obtained in \eqref{soln02},\eqref{soln03} and \eqref{soln04}.

\subsection{A Different Choice}

So far in this section, we only focused on the Cartan planes related to the momentum direction of the particle on the celestial $S^{p-1} \times S^{q-1}$. The rest of the Cartans we viewed as rotation generators that do not mix the signatures because what we did with those directions did not affect our understanding of the Mellin transform. 

However, it turns out that the choice of our generators can be made a bit more elegant if the Abelian subset comprising of the Cartan generators is chosen in such a way that it consists of the maximum number of boost-type generators. This number is $2 \times \min(m,n)$. The rest of the generators, which are $|m-n|$ in number, have to necessarily be of rotation-type. We will present some details of this choice in this subsection, even though the choices in the rest of the directions do not affect the general features of our discussion.  In particular, we have checked that things like Mellin transform, the relation between $L$ and $H$ primaries, etc., lead to parallel expressions, so we will not present those.

We arrange the coordinates as
\begin{equation}
	X = \{x_{1}, y_{1}, x_{2}, y_{2}.\dots,x_{2n}, y_{2n}, y_{2n+1},\dots,y_{2m} \}
\end{equation}
and write the Cartan generators and ladder operators by picking coordinates from the two coordinate ranges $x_{1}, y_{1}, x_{2}, y_{2}.\dots,x_{2n}, y_{2n}$ and $y_{2n+1},\dots,y_{2m}$. This way we have $2$ types of Cartan generators and $3$ types of raising and lowering operators. Specifically, picking both coordinates from $x_{1}, y_{1}, x_{2}, y_{2}.\dots,x_{2n}, y_{2n}$ give us the Cartan generator \eqref{cart1} and the ladder operators \eqref{raise1} and \eqref{lower1}. Picking both coordinates from $y_{2n+1},\dots,y_{2m}$ give the Cartan generators \eqref{cart2} and the ladder operators \eqref{raise3} and \eqref{lower3}. Lastly, picking one coordinate from$x_{1}, y_{1}, x_{2}, y_{2}.\dots,x_{2n}, y_{2n}$ and another from $y_{2n+1},\dots,y_{2m}$ give us the ladder operators \eqref{raise2} and \eqref{lower2}. Since Cartan generators are picked by pairing adjacent coordinates, there are no Cartan generators consisting of coordinates from $x_{1}, y_{1}, x_{2}, y_{2}.\dots,x_{2n}, y_{2n}$ and $y_{2n+1},\dots,y_{2m}$ both.

We write the explicit Cartan generators as
\begin{align}
	\t{H}_{I} &= i M_{I \t{I}} \;\; \forall \;\; I \in 1,\dots, 2n \label{cart1} \\
	\t{H}_{\t{j}} &=  -M_{2\t{j}-1, 2\t{j}} \;\; \forall \; \t{j} \in n+1,\dots, m \label{cart2}
\end{align}
Corresponding to these Cartan generators, we have the raising operators
\begin{align}
	\t{E}^{(1)}_{e^{J}\pm e^{K}} &=  \frac{1}{2}\{ (M_{J K} \pm M_{\t{J},\t{K}}) + (M_{K \t{J}} \mp M_{J \t{K} }) \} \;\; \forall \; K > J \in 1,\dots,2n \label{raise1}\\ 
	\t{E}^{(2)}_{e^{J} \pm e^{\t{ j } } }&= -\frac{i}{2}\{ (M_{J, 2\t{j}-1} - M_{\t{J}, 2\t{j}-1 } ) \mp i(M_{J, 2\t{j} } - M_{\t{J}, 2\t{j} } )\} \forall J \in 1,\dots,2n \; \& \;  \t{j} \in n+1,\dots,m \label{raise2} \\
	\t{E}^{(3)}_{e^{\t{j}} \pm e^{\t{k}} } &= -\frac{1}{2}\{(M_{2\t{j} -1, 2\t{k} -1 }  - i M_{2\t{j} , 2\t{k} -1 } ) \mp i(M_{2\t{j} -1, 2\t{k}}  - i M_{2\t{j} , 2\t{k}} ) \} \forall \t{k} > \t{j} \in n+1,\dots,m, \label{raise3}
\end{align}
and the lowering operators
\begin{align}
	\t{F}^{(1)}_{e^{J}\pm e^{K}} &=   \frac{1}{2}\{ (M_{J K} \pm M_{\t{J},\t{K}}) - (M_{K \t{J}} \mp M_{J \t{K} }) \} \;\; \forall \; K > J \in 1,\dots,2n \label{lower1} \\ 
	\t{F}^{(2)}_{e^{J} \pm e^{\t{ j } } }&=  \frac{i}{2}\{ (M_{J, 2\t{j}-1} + M_{\t{J}, 2\t{j}-1 } ) \pm i(M_{J, 2\t{j} } + M_{\t{J}, 2\t{j} } )\} \forall J \in 1,\dots,2n \; \& \;  \t{j} \in n+1,\dots,m \label{lower2} \\
	\t{F}^{(3)}_{e^{\t{j}} \pm e^{\t{k}} } &= \frac{1}{2}\{(M_{2\t{j} -1, 2\t{k} -1 }  + i M_{2\t{j} , 2\t{k} -1 } ) \pm i(M_{2\t{j} -1, 2\t{k}}  + i M_{2\t{j} , 2\t{k}} ) \} \forall \t{k} > \t{j} \in n+1,\dots,m. \label{lower3}
\end{align}
A simple counting of the total number of generators (Cartan generators and the ladder operators) gives the expected value of $(n+m)(2(n+m)-1)$.\par 
The relevant set of commutation relations satisfied by \eqref{cart1} - \eqref{lower3} are as follows
\begin{align}
	[\t{H}_{I}, \t{E}^{(1)}_{e^{J} \pm e^{K}}] &= -(\eta_{I J} \pm \eta_{I K}) \t{E}^{(1)}_{e^{J} \pm e^{K}}\label{commute1}\\
	[\t{H}_{I}, \t{F}^{(1)}_{e^{J} \pm e^{K}}] &= (\eta_{I J} \pm \eta_{I K}) \t{F}^{(1)}_{e^{J} \pm e^{K}}\label{commute2}\\
	[\t{E}^{(1)}_{e^{J} \pm e^{K}}, \t{F}^{(1)}_{e^{J} \pm e^{K}}] &= -(\eta_{K K} \t{H}_{J} \pm \eta_{J J}\t{H}_{K})\label{commute3}\\
	[\t{H}_{I}, \t{E}^{(2)}_{e^{J} \pm e^{\t{j}}}] &= -\eta_{I J}\t{E}^{(2)}_{e^{J} \pm e^{\t{j}}}\label{commute4}\\
	[\t{H}_{\t{k}}, \t{E}^{(2)}_{e^{J} \pm e^{\t{j}}}] &= \mp \eta_{2\t{j}, 2\t{k}}\t{E}^{(2)}_{e^{J} \pm e^{\t{j}}}\label{commute5}\\
	[\t{H}_{I}, \t{F}^{(2)}_{e^{J} \pm e^{\t{j}}}] &= \eta_{I J}\t{F}^{(2)}_{e^{J} \pm e^{\t{j}}}\label{commute6}\\
	[\t{H}_{\t{k}}, \t{F}^{(2)}_{e^{J} \pm e^{\t{j}}}] &= \pm \eta_{2\t{j}, 2\t{k}}\t{F}^{(2)}_{e^{J} \pm e^{\t{j}}}\label{commute7}\\
	[\t{E}^{(2)}_{e^{J} \pm e^{\t{j} }}, \t{F}^{(2)}_{e^{J} \pm e^{\t{j} }}] &= -(\eta_{2\t{j}, 2\t{j}}\t{H}_{J} \pm \eta_{J,J}\t{H}_{\t{j}})\label{commute8}\\
	[\t{H}_{\t{i}}, \t{E}^{(3)}_{e^{\t{j}} \pm e^{\t{k}}}] &= -(\eta_{2\t{i},2\t{j}} \pm \eta_{2\t{i},2\t{k}})\t{E}^{(3)}_{e^{\t{j}} \pm e^{\t{k}}} \label{commute9}\\
	[\t{H}_{\t{i}}, \t{F}^{(3)}_{e^{\t{j}} \pm e^{\t{k}}}] &= (\eta_{2\t{i},2\t{j}} \pm \eta_{2\t{i},2\t{k}})\t{F}^{(3)}_{e^{\t{j}} \pm e^{\t{k}}} \label{commute10}\\
	[\t{E}^{(3)}_{e^{\t{j}} \pm e^{\t{k}}}, \t{F}^{(3)}_{e^{\t{j}} \pm e^{\t{k}}}] &= -(\eta_{2\t{k}, 2\t{k}}\t{H}_{\t{j}} \pm \eta_{2\t{j}, 2\t{j}}\t{H}_{\t{k}}) \label{commute11}
\end{align}
Clearly the generators and algebra can be written more compactly, with these choices than in the previous subsections. 
With these commutation relations, we can proceed with solving for the highest weight state $ \Phi_{h}$. It is given by the solution of the following differential equations
\begin{align}
	\t{H}_{I}\Phi_{h} &= \t{h}_{I}\Phi_{h} \label{high1}\\
	\t{H}_{\t{j}}\Phi_{h} &= \t{h}_{\t{j}}\Phi_{h} \label{high2}\\
	\t{E}^{(1)}_{e^{J} \pm e^{K}}\Phi_{h} &= 0 \label{high3}\\
	\t{E}^{(2)}_{e^{J} \pm e^{\t{j} } }\Phi_{h} &=0 \label{high4}\\
 \t{E}^{(3)}_{e^{\t{j} } \pm e^{\t{k} } }\Phi_{h} &= 0 \label{high5}
\end{align}
From this point on the discussion proceeds quite parallelly with the previous subsections, as long as sufficient attention is paid to the various indices and their ranges.

\section{The $SO(2,2)$ Comparison}

We will compare our explicit expressions in Section 4 and Section 5, specialized to the $SO(2,2)$ case, with the results in \cite{Strominger} where the results are presented in the $SL(2,R) \times SL(2,R)$ language. The possibility of translating $SO(2,2)$ to $SL(2,R) \times SL(2,R)$ is an accidental isomorphism at the level of the Lie algebras and is not available in higher dimensions. So it behooves us to check that the results match. We will find that the expressions match exactly.

In our language the coordinates in $SO(2,2)$ are arranged as
\begin{equation}
	X = \{x_{1},x_{2},y_{1},y_{2}\}
\end{equation}
And they are paramterized as follows (according to \eqref{1.10} )
\begin{align}
	x_{1} &= r \cos\theta \\
	x_{2} &= r \sin\theta \\
	y_{1} &= t \cos\phi \\
	y_{2}  &= t \sin\phi
\end{align}
The $L$-Primary generators \eqref{Klein-Cartan}-\eqref{KleinLowering2} reduce to the following generators when we restrict to $n=m=1$ in $SO(2n,2m)$. 
\begin{align}
	H_{1} &= M_{1 2} = -i \partial_{\theta} \\
	H_{\t{1}} &= M_{\t{1} \t{2}} = i \partial_{\phi} \\
	E_{e^{1}\pm e^{\t{1} }} &= \frac{i}{2}e^{i(\theta \pm \phi)}\{(r\partial_{t} + t\partial_{r}) \pm i \frac{r}{t}\partial_{\phi} + i\frac{t}{r}\partial_{\theta} \} \\
	F_{e^{1} \pm e^{\t{1} } } &= -\frac{i}{2}e^{-i(\theta \pm \phi)}\{(r\partial_{t} + t\partial_{r}) \mp i \frac{r}{t}\partial_{\phi} - i\frac{t}{r}\partial_{\theta} \}
\end{align}
Since the constraint on the coordinates is the same as \eqref{4.40pan}, we have the further parametrization of the $r, t$ coordinates as
\begin{align}
	r &= \tau \cosh \rho \\
	t &= \tau \sinh \rho
\end{align}
This casts the generators in the following form
\begin{align}
	H_{1} &= M_{1 2} = -i \partial_{\theta} \\
	H_{\t{1}} &= M_{\t{1} \t{2}} = i \partial_{\phi} \\
	E_{e^{1}\pm e^{\t{1} }} &= \frac{i}{2}e^{i(\theta \pm \phi)}\{\partial_{\rho} \pm i \coth\rho \partial_{\phi} + i\tanh\rho \partial_{\theta} \} \\
	F_{e^{1} \pm e^{\t{1} } } &= -\frac{i}{2}e^{-i(\theta \pm \phi)}\{\partial_{\rho} \mp i \coth\rho \partial_{\phi} - i\tanh\rho \partial_{\theta} \}
\end{align}
According to the notation in Equation(s) 3.2 of \cite{Strominger}, we then have the following relations
\begin{align}
	L_{1} &= i F_{e^{1} + e^{\t{1} }}\\
	L_{0} &= \frac{1}{2} (H_{1} - H_{\t{1}} )\\
	L_{-1} &= i E_{e^{1} + e^{\t{1} }}\\
	\bar{L}_{1} &= i F_{e^{1} - e^{\t{1 } }}\\
	\bar{L}_{0} &= \frac{1}{2}(H_{1} + H_{\t{1} })\\
	\bar{L}_{-1} &= i E_{e^{1} - e^{\t{1 } }}
\end{align} 
The left hand sides are variables defined in \cite{Strominger}, and the right hand sides follow our notation. 
Note that our coordinates map to the coordinates used in Equation 3.2 of \cite{Strominger} through 
\begin{align}
	\phi_{\text{this paper} } \rightarrow \phi_{\cite{Strominger} }, \ \  \rho_{\text{this paper} } \rightarrow \rho_{\cite{Strominger}}, \ \ \theta_{\text{this paper} } \rightarrow \psi_{\cite{Strominger}}.
\end{align}

We now compare the $H$-primary generators in \cite{Strominger}  with the $H$-primary generators we write for $SO(p,q)$, restricted to $p=q=2$. From \eqref{boostCart1}-\eqref{lower1min}, we can write the $SO(2,2)$ case as follows
\begin{align}
	\t{H}_{1} &= i M_{1 \t{1}} \label{boostCart1SO22}\\
	\t{H}_{2} &= i M_{2 \t{2}} \label{boostCart2SO22}
\end{align}
The ladder operators are the following
\begin{align}
	\t{E}^{(1)}_{e^{1}\pm e^{2}} &=  \frac{1}{2}\{ (M_{1 2} \pm M_{\t{1},\t{2}}) + (M_{2 \t{1}} \mp M_{1 \t{2} }) \}  \label{raise1minSO22}\\ 
	\t{F}^{(1)}_{e^{1}\pm e^{2}} &=  \frac{1}{2}\{ (M_{1 2} \pm M_{\t{1},\t{2}}) - (M_{2 \t{1}} \mp M_{1 \t{2} }) \} \label{lower1minSO22}
\end{align}
To relate these to Equation (5.2) of \cite{Strominger}, we need to note that a choice in the ordering of coordinates has been made when deciding which boost generators we are picking. It turns out that this choice maps in the following way from our paper to \cite{Strominger}:
\begin{equation}
	\{x_{1}, y_{1}, x_{2}, y_{2}\}_{\text{Here} } \equiv \{x_{2}, y_{1}, x_{1}, y_{2}\}_{\text{There} }
\end{equation}
One sees that a precise map exists under the following identification of the Cartans
\begin{align}
	H^{\hat{x} }_{0} &= \frac{1}{2}(\t{H}_{1} + \t{H}_{\t{1} })\\
	\bar{H}^{\hat{x} }_{0} &= \frac{1}{2}(\t{H}_{1} -\t{H}_{\t{1} } )
\end{align}
and the ladder operators
\begin{align}
	H^{\hat{x} }_{+1} = - i \t{F}^{(1)}_{e^{1} + e^{2}} \\
	H^{\hat{x}}_{-1} = - i 	\t{E}^{(1)}_{e^{1} + e^{2}}\\
	\bar{H}^{\hat{x}}_{+1} = - i \t{F}^{(1)}_{e^{1} - e^{2}}\\
	\bar{H}^{\hat{x}}_{+1} = - i \t{E}^{(1)}_{e^{1} - e^{2}}
\end{align}
Again  the left hand sides are variables defined in \cite{Strominger}, and the right hand sides follow our notation. 

The ladder operators that we use and the ones used in \cite{Strominger} have a relative ``$i$'' between them, but the Cartan generators are (essentially) the same. This can be traced to the fact that our definition \eqref{Ladder-def} follows\footnote{Note that our definition of the ladder operators in terms of $M_{IJ}$ are identical to that in \cite{Georgi}, but our $M_{IJ}$ satisfy the Klenian signature algebra instead of the Euclidean algebra. So our Cartan-Weyl form \eqref{axx3}-\eqref{fullCommutator} is of course different from that of the Euclidean $so(p+q)$ algebra considered in \cite{Georgi}.} that of \cite{Georgi}, while the ones used in \cite{Strominger} effectively have an extra $i$ relative to those in \cite{Georgi}. Let us be a bit more explicit about this. The relative ``$i$" in \cite{Strominger} simply exchanges the raising and lowering operators in our definition vs theirs. This can be seen, eg., from the $SL(2,R)$ algebra (3.5) in \cite{Strominger}. Our definition adds an $i$ to $L_1$ and to $L_{-1}$, while there is no $i$ in $L_0$. This results in an extra sign in the $[L_1, L_{-1}]$ commutator, while the other two commutators remain intact. Therefore, because our raising and lowering operators are exchanged with respect to those in \cite{Strominger}, the sign is precisely taken care of.

Note also that the Ladder operator pieces in the quadratic Casimir in \cite{Strominger} have an extra negative sign relative to ours to accommodate for this. So the final results match exactly.

\section*{Acknowledgments}

We thank Tanay Pathak and Pratik Nandy for helpful discussions. BB is supported by the Ministry of Human Resources, Government of India through the Prime Minister's Research Fellowship.

\appendix

\section{Useful Forms for the Generators}
\noindent
By using the parametrization of the coordinates given by \eqref{1.3}-\eqref{1.10}, and the expression for the generators (\eqref{gen1}-\eqref{gen3}), we have the following explicit forms for the generators. The indices are $j,k \in 1,\dots,n\;\; \forall\; j<k$.
\begin{align*}
M_{2j-1,2k-1} &= -i \Big(\cos\theta_{j}\cos\theta_{k}(r_{j}\partial_{r_{k}} - r_{k}\partial_{r_{j}}) - \frac{r_{j}}{r_{k}}\cos\theta_{j}\sin\theta_{k}\partial_{\theta_{k}} + \frac{r_{k}}{r_{j}}\cos\theta_{k}\sin\theta_{j}\partial_{\theta_{j}}\Big)\\
M_{2j-1,2k} &= -i \Big(\cos\theta_{j}\sin\theta_{k}(r_{j}\partial_{r_{k}} - r_{k}\partial_{r_{j}}) + \frac{r_{j}}{r_{k}}\cos\theta_{j}\cos\theta_{k}\partial_{\theta_{k}} + \frac{r_{k}}{r_{j}}\sin\theta_{k}\sin\theta_{j}\partial_{\theta_{j}}\Big)\\
M_{2j,2k-1}&= -i \Big(\sin\theta_{j}\cos\theta_{k}(r_{j}\partial_{r_{k}} - r_{k}\partial_{r_{j}}) - \frac{r_{j}}{r_{k}}\sin\theta_{j}\sin\theta_{k}\partial_{\theta_{k}} - \frac{r_{k}}{r_{j}}\cos\theta_{k}\cos\theta_{j}\partial_{\theta_{j}}\Big)\\
M_{2j,2k}&= -i \Big(\sin\theta_{j}\sin\theta_{k}(r_{j}\partial_{r_{k}} - r_{k}\partial_{r_{j}}) + \frac{r_{j}}{r_{k}}\sin\theta_{j}\cos\theta_{k}\partial_{\theta_{k}} - \frac{r_{k}}{r_{j}}\sin\theta_{k}\cos\theta_{j}\partial_{\theta_{j}}\Big)
\end{align*}\label{pant1}\noindent
From these expressions, we can write down the explicit forms for the raising operators as follows (the superscript  ``$x,x$'' is there to indicate that both $j$ and $k$ belong to the $x$ coordinates)
\begin{equation}\label{3.13}
	E^{x,x}_{e^{j}\pm e^{k}} = -\frac{i}{2} e^{i(\theta_{j}\pm \theta_{k})}\{(r_{j}\partial_{r_{k}} - r_{k}\partial_{r_{j}}) \pm i\frac{r_{j}}{r_{k}}\partial_{\theta_{k}} - i\frac{r_{k}}{r_{j}}\partial_{\theta_{j}}\}
\end{equation}
The lowering operators are 
\begin{equation}\label{3.14}
	F^{x,x}_{e^{j}\pm e^{k}} = \frac{i}{2} e^{-i(\theta_{j}\pm \theta_{k})}\{(r_{j}\partial_{r_{k}} - r_{k}\partial_{r_{j}}) \mp i\frac{r_{j}}{r_{k}}\partial_{\theta_{k}} + i\frac{r_{k}}{r_{j}}\partial_{\theta_{j}}\}
\end{equation}
Similarly, for the case where $n+\tilde{j},n+\tilde{k} \in n+1,\dots,n+m\;\; \forall\; \tilde{j}<\tilde{k}$, we have
\begin{align*}\label{pant2} 
	M_{2\tilde{j}-1,2\tilde{k}-1} &= i \Big(\cos\phi_{\tilde{j}}\cos\phi_{\tilde{k}}(t_{\tilde{j}}\partial_{t_{\tilde{k}}} - t_{\tilde{k}}\partial_{t_{\tilde{j}}}) - \frac{t_{\tilde{j}}}{t_{\tilde{k}}}\cos\phi_{\tilde{j}}\sin\phi_{\tilde{k}}\partial_{\phi_{\tilde{k}}} + \frac{t_{\tilde{k}}}{t_{\tilde{j}}}\cos\phi_{\tilde{k}}\sin\phi_{\tilde{j}}\partial_{\phi_{\tilde{j}}}\Big)\\
	M_{2\tilde{j}-1,2\tilde{k}} &= i \Big(\cos\phi_{\tilde{j}}\sin\phi_{\tilde{k}}(t_{\tilde{j}}\partial_{t_{\tilde{k}}} - t_{\tilde{k}}\partial_{t_{\tilde{j}}}+ \frac{t_{\tilde{j}}}{t_{\tilde{k}}}\cos\phi_{\tilde{j}}\cos\phi_{\tilde{k}}\partial_{\phi_{\tilde{k}}} + \frac{t_{\tilde{k}}}{t_{\tilde{j}}}\sin\phi_{\tilde{k}}\sin\phi_{\tilde{j}}\partial_{\phi_{\tilde{j}}}\Big)\\
	M_{2\tilde{j},2\tilde{k}-1}&= i \Big(\sin\phi_{\tilde{j}}\cos\phi_{\tilde{k}}(t_{\tilde{j}}\partial_{t_{\tilde{k}}} - t_{\tilde{k}}\partial_{t_{\tilde{j}}}) - \frac{t_{\tilde{j}}}{t_{\tilde{k}}}\sin\phi_{\tilde{j}}\sin\phi_{\tilde{k}}\partial_{\phi_{\tilde{k}}} - \frac{t_{\tilde{k}}}{t_{\tilde{j}}}\cos\phi_{\tilde{k}}\cos\phi_{\tilde{j}}\partial_{\phi_{\tilde{j}}})\\
	M_{2\tilde{j},2\tilde{k}}&= i \Big(\sin\phi_{\tilde{j}}\sin\phi_{\tilde{k}}(t_{\tilde{j}}\partial_{t_{\tilde{k}}} - t_{\tilde{k}}\partial_{t_{\tilde{j}}}) + \frac{t_{\tilde{j}}}{t_{\tilde{k}}}\sin\phi_{\tilde{j}}\cos\phi_{\tilde{k}}\partial_{\phi_{\tilde{k}}} - \frac{t_{\tilde{k}}}{t_{\tilde{j}}}\sin\phi_{\tilde{k}}\cos\phi_{\tilde{j}}\partial_{\phi_{\tilde{j}}}\Big)
\end{align*}
Likewise, in this case we have the following raising operators
\begin{equation}\label{3.15}
	E^{y,y}_{e^{\tilde{j}}\pm e^{\tilde{k}}} = \frac{i}{2} e^{i(\phi_{\tilde{j}}\pm \phi_{\tilde{k}})}\{(t_{\tilde{j}}\partial_{t_{\tilde{k}}} - t_{\tilde{k}}\partial_{t_{\tilde{j}}}) \pm i\frac{t_{\tilde{j}}}{t_{\tilde{k}}}\partial_{\phi_{\tilde{k}}} - i\frac{t_{\tilde{k}}}{t_{\tilde{j}}}\partial_{\phi_{\tilde{j}}}\}
\end{equation}
and lowering operators
\begin{equation}\label{3.16}
	F^{y,y}_{e^{\tilde{j}}\pm e^{\tilde{k}}} = -\frac{i}{2} e^{-i(\phi_{\tilde{j}}\pm \phi_{\tilde{k}})}\{(t_{\tilde{j}}\partial_{t_{\tilde{k}}} - t_{\tilde{k}}\partial_{t_{\tilde{j}}}) \mp i\frac{t_{\tilde{j}}}{t_{\tilde{k}}}\partial_{\phi_{\tilde{k}}} + i\frac{t_{\tilde{k}}}{t_{\tilde{j}}}\partial_{\phi_{\tilde{j}}}\}
\end{equation}
Finally, we look at the case where $j \in 1,\dots,n ; \;\;\tilde{j}+n \in n+1,\dots,n+m$.
\begin{align*}
	M_{2j-1,2\tilde{j}-1} &= i\Big(\cos\theta_{j}\cos\phi_{\tilde{j}}(r_{j}\partial_{t_{\tilde{j}}} + t_{\tilde{j}}\partial_{r_{j}}) - \frac{r_{j}}{t_{\tilde{j}}}\sin\phi_{\tilde{j}}\cos\theta_{j}\partial_{\phi_{\tilde{j}}} - \frac{t_{\tilde{j}}}{r_{j}}\cos\phi_{\tilde{j}}\sin\theta_{j}\partial_{\theta_{j}}\Big)\\
	M_{2j-1,2\tilde{j}} &= i\Big(\cos\theta_{j}\sin\phi_{\tilde{j}}(r_{j}\partial_{t_{\tilde{j}}} + t_{\tilde{j}}\partial_{r_{j}}) + \frac{r_{j}}{t_{\tilde{j}}}\cos\phi_{\tilde{j}}\cos\theta_{j}\partial_{\phi_{\tilde{j}}} - \frac{t_{\tilde{j}}}{r_{j}}\sin\phi_{\tilde{j}}\sin\theta_{j}\partial_{\theta_{j}}\Big)\\
	M_{2j,2\tilde{j}-1} &= i\Big(\sin\theta_{j}\cos\phi_{\tilde{j}}(r_{j}\partial_{t_{\tilde{j}}} + t_{\tilde{j}}\partial_{r_{j}}) - \frac{r_{j}}{t_{\tilde{j}}}\sin\phi_{\tilde{j}}\sin\theta_{j}\partial_{\phi_{\tilde{j}}} + \frac{t_{\tilde{j}}}{r_{j}}\cos\phi_{\tilde{j}}\cos\theta_{j}\partial_{\theta_{j}}\Big)\\
	M_{2j,2\tilde{j}} &= i\Big(\sin\theta_{j}\sin\phi_{\tilde{j}}(r_{j}\partial_{t_{\tilde{j}}} + t_{\tilde{j}}\partial_{r_{j}}) + \frac{r_{j}}{t_{\tilde{j}}}\cos\phi_{\tilde{j}}\sin\theta_{j}\partial_{\phi_{\tilde{j}}} + \frac{t_{\tilde{j}}}{r_{j}}\sin\phi_{\tilde{j}}\cos\theta_{j}\partial_{\theta_{j}}\Big)
\end{align*}\label{pant3}
Here the raising operators take the form
\begin{equation}\label{3.17}
	E^{x,y}_{e^{j}\pm e^{\tilde{j}}} = \frac{i}{2}e^{i(\theta_{j}\pm \phi_{\tilde{j}})}\{(r_{j}\partial_{t_{\tilde{j}}} + t_{\tilde{j}}\partial_{r_{j}}) \pm i \frac{r_{j}}{t_{\tilde{j}}}\partial_{\phi_{\tilde{j}}} + i\frac{t_{\tilde{j}}}{r_{j}}\partial_{\theta_{j}}\}
\end{equation}
and the lowering operators 
\begin{equation}\label{3.18}
	F^{x,y}_{e^{j}\pm e^{\tilde{j}}} = -\frac{i}{2}e^{-i(\theta_{j}\pm \phi_{\tilde{j}})}\{(r_{j}\partial_{t_{\tilde{j}}} + t_{\tilde{j}}\partial_{r_{j}}) \mp i \frac{r_{j}}{t_{\tilde{j}}}\partial_{\phi_{\tilde{j}}} - i\frac{t_{\tilde{j}}}{r_{j}}\partial_{\theta_{j}}\}
\end{equation}
Now, we turn to the Cartan generators. These can be written as 
\begin{align}
	H_{j} = M_{2j-1,2j} &= -i\partial_{\theta_{j}}\\
	H_{\tilde{j}} = M_{2\tilde{j} -1,2\tilde{j}} &= i\partial_{\phi_{\tilde{j}}}
\end{align}
\section{Simplifying the Weight Conditions}

\subsection{Highest Weight}
\noindent
We can write the set of equations that must be satisfied by the highest weight state $\Phi_{h}$. These are as follows
\begin{align}
	H_{j}\Phi_{++} &= \lambda_{i}\Phi_{++} \label{eigen1} \;\; \forall j \\
	H_{\tilde{j}}\Phi_{++} &= \tilde{\lambda}_{\tilde{j}}\Phi_{++} \label{eigen2} \;\; \forall \t{j} \\
	E^{x,x}_{e^{j}\pm e^{k}}\Phi_{++} &= E^{y,y}_{e^{\tilde{j}}\pm e^{\tilde{k}}}\Phi_{++} = 0 \;\; \forall \; k > j \; \& \; \t{k} > \t{j} \\ E^{x,y}_{e^{j}\pm e^{\tilde{j}}}\Phi_{++} &= 0 \;\; \forall \; j, \t{j}
\end{align}
A simple sanity check before we begin, is to note that the the conditions on the last two lines add up to $n(n-1)+m(m-1)+2nm$ which is $(n+m)(n+m-1)$ conditions. This is half the total number of roots of $SO(2n,2m)$. The other half of the roots are covered by the lowering operators.
We can unpack these equations one by one to arrive at the final set of equations satisfied by $\Phi_{++}$. Let us start with
\begin{equation}
	\begin{split}
		&E^{x,x}_{e^{j}\pm e^{k}}\Phi_{++} = 0\\
		\implies& (r_{j}\partial_{r_{k}} - r_{k}\partial_{r_{j}})\Phi_{++} \mp \frac{r_{j}}{r_{k}}\lambda_{k}\Phi_{++} + \frac{r_{k}}{r_{j}}\lambda_{j}\Phi_{++} = 0
	\end{split}\label{4.13pa}
\end{equation}
Adding and subtracting the two equations in \eqref{4.13pa} gives
\begin{align}
	(r_{j}\partial_{r_{k}} - r_{k}\partial_{r_{j}})\Phi_{++}  + \frac{r_{k}}{r_{j}}\lambda_{j}\Phi_{++} &= 0\\
	\frac{r_{j}}{r_{k}}\lambda_{k}\Phi_{++} &= 0
\end{align}
From the second equation above, we can conclude that $\lambda_{k} = 0 \; \forall k>1$. The $k>1$ condition arises because $k>j$. We effectively have the following expressions (which need to be explicitly solved)
\begin{align}
	(r_{1}\partial_{r_{k}} - r_{k}\partial_{r_{1}})\Phi_{++}  + \frac{r_{k}}{r_{1}}\lambda_{1}\Phi_{++} &= 0 \; \forall k > 1\\
	(r_{j}\partial_{r_{k}} - r_{k}\partial_{r_{j}})\Phi_{++} &= 0 \; \forall j,k > 1 \; \text{and} \;k>j
\end{align}
Similarly, the action of the $E^{y,y}$ raising operators can be cast into the following form
\begin{align}
	(t_{1}\partial_{t_{\tilde{k}}} - t_{\tilde{k}}\partial_{t_{1}})\Phi_{++}  - \frac{t_{\tilde{k}}}{t_{1}}\tilde{\lambda}_{1}\Phi_{++} &= 0 \; \forall \tilde{k} > 1 \label{4.18pan}\\
	(t_{\tilde{j}}\partial_{t_{\tilde{k}}} - t_{\tilde{k}}\partial_{t_{\tilde{j}}})\Phi_{++} &= 0 \; \forall \tilde{j},\tilde{k} > 1 \; \text{and} \;\tilde{k}>\tilde{j} \label{4.19pan}
\end{align}
where all $\tilde{\lambda}_{\tilde{k}} = 0 \; \forall \tilde{k}>1$.\par 
Finally, we turn to the expressions we obtain from the $E^{x,y}$ type raising operators. Followig the same steps as before, one can arrive at the conclusion that $\tilde{\lambda}_{\tilde{j}} = 0 \; \forall \tilde{j}$. The equations (arising in this case) satisfied by $\Phi_{++}$ are given as follows
\begin{align}
	(r_{1}\partial_{t_{\tilde{j}}} + t_{\tilde{j}}\partial_{r_{1}})\Phi_{++} - \frac{t_{\tilde{j}}}{r_{1}}\lambda_{1}\Phi_{++} &= 0\; \forall \tilde{j} \\
	(r_{j}\partial_{t_{\tilde{j}}} + t_{\tilde{j}}\partial_{r_{j}})\Phi_{++} &= 0\; \forall \tilde{j} \; \text{and} \; j> 1
\end{align}
Note that since we have $\tilde{\lambda}_{1} = 0$, \eqref{4.18pan} reduces to \eqref{4.19pan}. Therefore, the final set of equations satisfied by $\Phi_{++}$ are as follows 
\begin{align}
	\frac{1}{r_{k}}\partial_{r_k}\Phi_{++} &= \frac{1}{r_{1}}\partial_{r_1}\Phi_{++} - \frac{\lambda_{1}}{r_{1}^{2}}\Phi_{++}\;\; \forall\;\; k \neq 1 \label{4.35pan}\\
	-\frac{1}{t_{\t{j} }}\partial_{t_{\t{j} }}\Phi_{++}&= \frac{1}{r_{1}}\partial_{r_1}\Phi_{++} - \frac{\lambda_{1}}{r_{1}^{2}}\Phi_{++}\;\; \forall\;\; \t{j} \label{4.36pan}\\
	\frac{1}{r_{k}}\partial_{r_k}\Phi_{++} &= \frac{1}{r_{j}}\partial_{r_j}\Phi_{++}\;\; \forall \;\; k\;,\;j \neq 1 \label{4.37pan}\\
	\frac{1}{t_{\t{j} }}\partial_{t_{\t{j} } }\Phi_{++} &= \frac{1}{t_{\t{k}}}\partial_{t_{\t{k}}}\Phi_{++} \;\; \forall \;\; \t{j}, \t{k}\label{4.38pan}\\
	\frac{1}{r_{k}}\partial_{r_k}\Phi_{++} &= - \frac{1}{t_{\t{j}}}\partial_{t_{\t{j}}}\Phi_{++} \;\; \forall \;\; \t{j}, \;\; \text{and} \;\; k \neq 1 \label{4.39pan}
\end{align}
The expressions \eqref{4.37pan}-\eqref{4.39pan} can be also obtained from \eqref{4.35pan} and \eqref{4.36pan}. In further discussions, we will only mention the equations similar to \eqref{4.35pan} and \eqref{4.36pan}. We will later solve for $\Phi_{++}$ (or the state that we are considering), subject to the following constraint
\begin{equation}\label{4.40pan}
	\sum_{\tilde{k}=1}^{n}t_{\tilde{k}}^{2} - \sum_{j=1}^{m}r_{j}^{2} = - \tau^{2}
\end{equation}

\subsection{Lowest Weight and Mixed Weight}
\noindent
The annihilation condition satisfied by the lowest weight state $\Phi_{--}$ are as follows
\begin{equation}
	F^{x,x}_{e^{j}\pm e^{k}}\Phi_{--} = F^{y,y}_{e^{\tilde{j}}\pm e^{\tilde{k}}}\Phi_{--} = F^{x,y}_{e^{j}\pm e^{\tilde{j}}}\Phi_{--} = 0 \label{annLOW1} \;\; \forall \; k > j \; \& \; \t{k} > \t{j}
\end{equation}

\noindent
The eigenvalue equations are given by \eqref{eigen1}-\eqref{eigen2}, with $\Phi_{++}$ replaced by $\Phi_{--}$. These along with the annihilation conditions \eqref{annLOW1} can be evaluated to give
\begin{align}
	\frac{1}{r_{k}}\partial_{r_k}\Phi_{--} &= \frac{1}{r_1}\partial_{r_1}\Phi_{--} + \frac{\lambda_{1}}{r_{1}^{2}}\Phi_{--}\;\; \forall\;\; k \neq 1 \label{4.35pan2}\\
	-\frac{1}{t_{\t{j}}}\partial_{t_{\t{j} }}\Phi_{--}&= \frac{1}{r_{1}}\partial_{r_1}\Phi_{--} + \frac{\lambda_{1}}{r_{1}^{2}}\Phi_{--}\;\; \forall\;\; \t{j} \label{4.36pan2}
\end{align}
The analogue of \eqref{4.37pan}-\eqref{4.39pan} for $\Phi_{--}$ follows from \eqref{4.35pan2}-\eqref{4.36pan2}.\par 
The mixed weight state of the first kind $\Phi_{+-}$ corresponds to the following annihilation conditions along with the eigenvalue equations \eqref{eigen1}-\eqref{eigen2} for $\Phi_{+-}$.
\begin{align}
	E^{x,x}_{e^{j} + e^{k}}\Phi_{+-} &= E^{y,y}_{e^{\tilde{j}} + e^{\tilde{k}}}\Phi_{+-} =E^{x,y}_{e^{j}+ e^{\tilde{j}}}\Phi_{+-} = 0  \label{annMIX1.1} \;\; \forall k > j \; \& \; \t{k} > \t{j}\\ 
	F^{x,x}_{e^{j} - e^{k}}\Phi_{+-} &= F^{y,y}_{e^{\tilde{j}} - e^{\tilde{k}}}\Phi_{+-} = F^{x,y}_{e^{j}-e^{\tilde{j}}}\Phi_{+-} = 0\label{annMIX1.2} \;\; \forall k > j \; \& \; \t{k} > \t{j}
\end{align}

\noindent
These can again be solved to give the following set of differential equations satisfied by $\Phi_{+-}$
\begin{align}
	-\frac{1}{r_{j}}\partial_{r_j}\Phi_{+-} &= \frac{1}{t_{m}}\partial_{t_m}\Phi_{+-} + \frac{\tilde{\lambda}_{m}}{t_{m}^{2}}\Phi_{+-}\;\; \forall\;\; j \label{4.35pan3}\\
	\frac{1}{t_{\t{j} } }\partial_{t_{\t{j} } }\Phi_{+-}&=  \frac{1}{t_{m}}\partial_{t_m}\Phi_{+-} + \frac{\tilde{\lambda}_{m}}{t_{m}^{2}}\Phi_{+-}\;\; \forall\;\; \t{j} < m\label{4.36pan3}
\end{align}
Lastly, the mixed weight state of the second kind $\Phi_{-+}$ corresponds to the following annihilation conditions along with the eigenvalue equations \eqref{eigen1}-\eqref{eigen2} for $\Phi_{+-}$ .
\begin{align}
	E^{x,x}_{e^{j} - e^{k}}\Phi_{-+} &= E^{y,y}_{e^{\tilde{j}} - e^{\tilde{k}}}\Phi_{-+}  = E^{x,y}_{e^{j}- e^{\tilde{j}}}\Phi_{-+} = 0 \label{annMIX2.1} \;\; \forall k > j \; \& \; \t{k} > \t{j}\\ 
	F^{x,x}_{e^{j} + e^{k}}\Phi_{-+} &= F^{y,y}_{e^{\tilde{j}} + e^{\tilde{k}}}\Phi_{-+} = F^{x,y}_{e^{j}+e^{\tilde{j}}}\Phi_{-+} = 0\label{annMIX2.2} \;\; \forall k > j \; \& \; \t{k} > \t{j}
\end{align}

\noindent
These yield
\begin{align}
	-\frac{1}{r_{j}}\partial_{r_j}\Phi_{-+} &= \frac{1}{t_{m}}\partial_{t_m}\Phi_{-+} - \frac{\tilde{\lambda}_{m}}{t_{m}^{2}}\Phi_{+-}\;\; \forall\;\; j \label{4.35pan4}\\
	\frac{1}{t_{\t{j} }}\partial_{t_{\t{j} } }\Phi_{-+}&=  \frac{1}{t_{m}}\partial_{t_m}\Phi_{-+} - \frac{\tilde{\lambda}_{m}}{t_{m}^{2}}\Phi_{-+}\;\; \forall\;\; \t{j} < m\label{4.36pan4}
\end{align}

\section{Primary State Solution}

\noindent
We will work with highest weight states in this section. The other weight cases are solved similarly. The equations we wish to solve are the following
\begin{align}\label{6.1t2}
	\frac{1}{r_{k}}\partial_{k}\Phi_{++} &= \frac{1}{r_{1}}\partial_{1}\Phi_{++} - \frac{h}{r_{1}^{2}}\Phi_{++}\;\; \forall\;\; k \neq 1\\
	-\frac{1}{t_{\tilde{j}}}\partial_{\tilde{j}}\Phi_{++}&= \frac{1}{r_{1}}\partial_{1}\Phi_{++} - \frac{h}{r_{1}^{2}}\Phi_{++}\;\; \forall\;\; \tilde{j} \label{6.2}
\end{align}
We will use the following parametrization\footnote{Let us emphasize that the $\theta$'s and $\phi$'s that we are introducing here should not be confused with the $\theta$'s and $\phi$'s introduced in section 2.1. There are only so many letters in the alphabet that can be intuitively used -- the $\theta$'s and $\phi$'s we use in this Appendix stays entirely within this Appendix. The reader if he/she wishes can just take the final result, the second expression in \eqref{finalA} from this Appendix, which is all that is needed in the main text.} 
\begin{equation}\label{6.3pan}
	\begin{split}
		r_{1} &= \tau \cosh \rho \cos \theta_{1} \\
		r_{2} &= \tau \cosh \rho \sin \theta_{1} \cos \theta_{2}\\
		&\vdots\\
		r_{n-1}& = \tau \cosh \rho \sin \theta_{1} \cdots \sin \theta_{n-2} \cos \theta_{n-1}\\
		r_{n} &= \tau \cosh \rho \sin \theta_{1} \cdots \sin \theta_{n-2} \sin \theta_{n-1}
	\end{split}
\end{equation}
\begin{equation}\label{6.4pan}
	\begin{split}
		t_{1} &= \tau \sinh \rho \cos \phi_{1} \\
		t_{2} &= \tau \sinh \rho \sin \phi_{1} \cos \phi_{2}\\
		&\vdots\\
		t_{m-1}& = \tau \sinh \rho \sin \phi_{1} \cdots \sin \phi_{m-2} \cos \phi_{m-1}\\
		t_{m} &= \tau \sinh \rho \sin \phi_{1} \cdots \sin \phi_{m-2} \sin \phi_{m-1}
	\end{split}
\end{equation}
Now, one can evaluate the equality $\frac{1}{r_{n}}\partial_{r_{n}}\Phi_{++} = \frac{1}{r_{n-1}}\partial_{r_{n-1}}\Phi_{++}$, and see what result we get. Using the relations given in \eqref{6.3pan} to \eqref{6.4pan}, we can write the following expressions for $\rho$ and $\theta,\phi$. These turn out to be 
\begin{align*}
	\tanh \rho &= \frac{\sqrt{t_{1}^{2}+t_{2}^{2} +\cdots+ t_{m}^{2}}}{\sqrt{r_{1}^{2} + r_{2}^{2} + \cdots + r_{n}^{2}}}\\
	\cot \theta_{i} &= \frac{r_{i}}{\sqrt{\sum_{j=i+1}^{n}r_{j}^{2}}}\\
	\cot \phi_{\tilde{j}} &= \frac{t_{\tilde{j}}}{\sqrt{\sum_{\tilde{k}=\tilde{j}+1}^{m}t_{j}^{2}}}
\end{align*}
From this one can evaluate $\frac{1}{r_{n}}\partial_{r_{n}}\Phi_{++}$ and $ \frac{1}{r_{n-1}}\partial_{r_{n-1}}\Phi_{++}$:
\begin{equation}\label{6.5pan}
	\frac{1}{r_{n}}\partial_{r_{n}}\Phi_{++} = -\frac{\tanh \rho}{\tau^{2}} \partial_{\rho}\Phi_{++} + \sum_{i=1}^{n-2}\frac{\cos^{2}\theta_{i}\cot \theta_{i}}{r_{i}^{2}}\partial_{\theta_{i}}\Phi_{++} + \frac{\cot \theta_{n-1}}{\tau^{2}\cosh^{2}\rho\sin^{2}\theta_{1}\cdots\sin^{2}\theta_{n-2}}\partial_{\theta_{n-1}}\Phi_{++}
\end{equation}
\begin{equation}\label{6.6pan}
	\frac{1}{r_{n-1}}\partial_{r_{n-1}}\Phi_{++} = -\frac{\tanh \rho}{\tau^{2}} \partial_{\rho}\Phi_{++} + \sum_{i=1}^{n-2}\frac{\cos^{2}\theta_{i}\cot \theta_{i}}{r_{i}^{2}}\partial_{\theta_{i}}\Phi_{++} - \frac{\tan \theta_{n-1}}{\tau^{2}\cosh^{2}\rho\sin^{2}\theta_{1}\cdots\sin^{2}\theta_{n-2}}\partial_{\theta_{n-1}}\Phi_{++}
\end{equation}
We can equate these two quantities because the right hand side of \eqref{6.1t2} is the same for all allowed values of $k$ including $n-1$ and $n$. We get
\begin{equation}\label{6.7pan}
	\partial_{\theta_{n-1}}\Phi_{++} = 0
\end{equation}
Now, we can apply \eqref{6.7pan} to \eqref{6.6pan} to get
\begin{equation}\label{6.8pan}
	\frac{1}{r_{n-1}}\partial_{r_{n-1}}\Phi_{++} = -\frac{\tanh \rho}{\tau^{2}} \partial_{\rho}\Phi_{++} + \sum_{i=1}^{n-2}\frac{\cos^{2}\theta_{i}\cot \theta_{i}}{r_{i}^{2}}\partial_{\theta_{i}}\Phi_{++}
\end{equation}
We can equate this to $\frac{1}{r_{n-2}}\partial_{r_{n-2}}\Phi_{++}$. For that we need to evaluate it first, and from that we get
\begin{equation}\label{6.9pan}
	\frac{1}{r_{n-2}}\partial_{r_{n-2}}\Phi_{++} = -\frac{\tanh \rho}{\tau^{2}} \partial_{\rho}\Phi_{++} + \sum_{i=1}^{n-3}\frac{\cos^{2}\theta_{i}\cot \theta_{i}}{r_{i}^{2}}\partial_{\theta_{i}}\Phi_{++} - \frac{\sin^{2}\theta_{n-2}\cot \theta_{n-2}}{r_{n-2}^{2}}\partial_{\theta_{n-2}}\Phi_{++}
\end{equation}
Comparing it to \eqref{6.8pan} and noting that the only different term as compared to \eqref{6.9pan} is the $\frac{\cos^{2}\theta_{n-2}\cot \theta_{n-2}}{r_{n-2}^{2}}\partial_{\theta_{n-2}}\Phi_{++}$ term, and thus equating \eqref{6.8pan} and \eqref{6.9pan}, we get
\begin{equation}\label{6.10pan}
	\partial_{\theta_{n-2}}\Phi_{++} = 0
\end{equation}
Following in the same way, we can get the results (by doing the same procedure for $\frac{1}{t_{\tilde{j}}}\partial_{t_{\tilde{j}}}\Phi_{++}$ as well)
\begin{equation}\label{6.11pan}
	\partial_{\theta_{i}}\Phi_{++} = 0\;\;\forall\;\; i \neq 1
\end{equation}
\begin{equation}\label{6.12pan}
	\partial_{\phi_{\tilde{j}}}\Phi_{++} = 0\;\;\forall\;\; \tilde{j}
\end{equation}
These, then allow us to simply consider the following equations
\begin{equation}\label{6.13pan}
	\begin{split}
		\frac{1}{r_{k}}\partial_{r_{k}}\Phi_{++} &= -\frac{\tanh \rho}{\tau^{2}} \partial_{\rho}\Phi_{++} + \frac{\cot \theta_{1}}{\tau^{2}\cosh^{2}\rho}\partial_{\theta_{1}}\Phi_{++}\\
		&=\frac{1}{r_{1}}\partial_{1}\Phi_{++} - \frac{h}{r_{1}^{2}}\Phi_{++}\\
		&=-\frac{\tanh \rho}{\tau^{2}} \partial_{\rho}\Phi_{++} - \frac{\tan \theta_{1}}{\tau^{2}\cosh^{2}\rho}\partial_{\theta_{1}}\Phi_{++} - \frac{h}{\tau^{2}\cosh^{2}\rho \cos^{2}\theta_{1}}\Phi_{++}
	\end{split}
\end{equation}
The first equality follows from explicit calculation together with the fact that all $\theta$-derivatives vanish. Second equality is just the right hand side of \eqref{6.1t2}, and the third equality again follows by direct calculation. 
Similarly, we also find the following equation
\begin{equation}\label{6.14pan}
	\begin{split}
		-\frac{1}{t_{\tilde{j}}}\partial_{t_{\tilde{j}}}\Phi_{++} &= -\frac{\coth \rho}{\tau^{2}}\partial_{\rho}\Phi_{++} = \frac{1}{r_{1}}\partial_{1}\Phi_{++} - \frac{h}{r_{1}^{2}}\Phi_{++} \\
		&=-\frac{\tanh \rho}{\tau^{2}} \partial_{\rho}\Phi_{++} - \frac{\tan \theta_{1}}{\tau^{2}\cosh^{2}\rho}\partial_{\theta_{1}}\Phi_{++} - \frac{h}{\tau^{2}\cosh^{2}\rho \cos^{2}\theta_{1}}\Phi_{++}
	\end{split}
\end{equation}
writing compactly, we have 
\begin{equation}\label{6.15pan}
	\frac{\cot \theta_{1}}{\tau^{2}\cosh^{2}\rho}\partial_{\theta_{1}}\Phi_{++} = - \frac{\tan \theta_{1}}{\tau^{2}\cosh^{2}\rho}\partial_{\theta_{1}}\Phi_{++} - \frac{h}{\tau^{2}\cosh^{2}\rho \cos^{2}\theta_{1}}\Phi_{++}
\end{equation}
\begin{equation}\label{6.16pan}
	-\frac{\coth \rho}{\tau^{2}}\partial_{\rho}\Phi_{++} = -\frac{\tanh \rho}{\tau^{2}} \partial_{\rho}\Phi_{++} - \frac{\tan \theta_{1}}{\tau^{2}\cosh^{2}\rho}\partial_{\theta_{1}}\Phi_{++} - \frac{h}{\tau^{2}\cosh^{2}\rho \cos^{2}\theta_{1}}\Phi_{++}
\end{equation}
These equations can be further combined to give the following
\begin{align}
	\partial_{\theta_{1}}\Phi_{++} = -h \tan\theta_{1} \Phi_{++}\label{finaleqn1}\\
	\partial_{\rho}\Phi_{++} = h \tanh\rho\ \Phi_{++}\label{finaleqn2}
\end{align}
Now, one can solve \eqref{finaleqn1} and \eqref{finaleqn2} to get the result by noting that the the general solutions from these two equations are respectively given as
\begin{align}
	\Phi_{++}(\rho, \theta_{1}) &= f_{1}(\rho)(\cos\theta_{1})^{h} + f_{2}(\rho) \;\;\; (\text{integrating} \; \eqref{finaleqn1}) \label{finalsoln1}\\
	\Phi_{++}(\rho, \theta_{1}) &= g_{1}(\theta_{1})(\cosh \rho)^{h} + g_{2}(\theta_{1}) \;\;\; (\text{integrating}\; \eqref{finaleqn2}) \label{finalsoln2}
\end{align}
where $f_{1}(\rho), f_{2}(\rho), g_{1}(\theta_{1})$ and $g_{2}(\theta_{1})$  are arbitrary functions of $\rho$ and $\theta_{1}$. Further, $f_{1}(\rho)$ and $g_{1}(\theta_{1})$ cannot be identically $0$ over the full range of $\rho$ and $\theta_{1}$ respectively. \par 
By equating \eqref{finalsoln1} and \eqref{finalsoln2}, we get the following result
\begin{equation}
	\Phi_{++} \propto (\cosh \rho \cos \theta_{1})^{h} = \Big(\frac{r_{1}}{\tau}\Big)^{h} \label{finalA}
\end{equation}
The reason for using the $\propto$ here is because we have not included the part of the solution that arises from solving the equations coming from the action of the Cartan generators on $\Phi_{2}$. With the solution to those equations included, we have the result \eqref{soln01}.

Similarly, the relevant equations can be solved for the other cases to obtain the results \eqref{soln02}, \eqref{soln03} and \eqref{soln04}.

\section{Quadratic Casimir in the $H$-Primary Basis}

We start by brielfy reviewing the well-known facts about the quadratic Casimir in a language that does not restrict itself to the compact form of the algebra. The Casimir is defined as
\begin{equation}
	c_{2} = g^{a b}t_{a}t_{b}
\end{equation}
where $t_{a}$,$t_{b}$ are the generators of the algebra, 
and $g^{ab}$ is defined via the struture constants 
\begin{equation}
	[t_{a},t_{b}] = C^{c}_{a b}t_{c}
\end{equation}
through
\begin{equation}
	g_{a b} \equiv C^{e}_{a d}C^{d}_{b c}.
\end{equation}
In our case, we have the following coefficients from the commutator of the generators $J_{A B}$:
\begin{equation}
	C^{ E F}_{A B,\; C D} = i(\eta_{A D}\delta^{E}_{B}\delta^{F}_{C} + \eta_{B C}\delta^{E}_{A}\delta^{F}_{D} -\eta_{A C}\delta^{E}_{B}\delta^{F}_{D} - \eta_{B D}\delta^{E}_{A}\delta^{F}_{C})
\end{equation}
Using this expression, we have
\begin{equation}
	\begin{split}
		g_{A B\; C D}  &= C^{E F}_{ A B,\; M N}C^{M N}_{ C D,\; E F}\\
		&= - \Big( \eta_{A N}\delta^{E}_{B}\delta^{F}_{M} + \eta_{B M}\delta^{E}_{A}\delta^{F}_{N} -\eta_{A M}\delta^{E}_{B}\delta^{F}_{N} - \eta_{B N}\delta^{E}_{A}\delta^{F}_{M}\Big)\\
		&\times \Big( \eta_{C F}\delta^{M}_{D}\delta^{N}_{E} + \eta_{D E}\delta^{M}_{C}\delta^{N}_{F} -\eta_{C E}\delta^{M}_{D}\delta^{N}_{F} - \eta_{B F}\delta^{M}_{C}\delta^{N}_{E}\Big)
	\end{split}
\end{equation}
From this, one see that the quadratic Casimir, upto multiplicative constants, is 
\begin{equation}
	c_{2} = \eta^{ A C}\eta^{ B D}J_{A B}J_{C D} \equiv J_{A B}J^{A B}
\end{equation}
One can of course also check that this is indeed the casimir, by observing that
\begin{equation}
	\eta^{ A B}\eta^{ B D}[J_{A B}J_{C D}, J_{M N}] = 0 \;\;\;\;\;\;\;\;\; \forall\;\;\; M,\;N \in \;{1,\cdots,2m+2n}
\end{equation}

\noindent
In the $H$-primary basis, the Casimir takes the form
\begin{equation}
	\begin{split}
		\nabla &= \sum_{I = 1}^{2}\t{H}_{I}\t{H}^{I} + \sum_{j=2}^{n}\t{H}_{j}\t{H}^{j} +\sum_{\t{j}=2}^{m}\t{H}_{\t{j}}\t{H}^{\t{j}}\\ &+   E^{(1)}_{e^{1} + e^{2}}F_{(1)}^{e^{1} + e^{2}} + F^{(1)}_{e^{1} + e^{2}}E_{(1)}^{e^{1} + e^{2}} 
		+  E^{(1)}_{e^{1} - e^{2}}F_{(1)}^{e^{1} - e^{2}} + F^{(1)}_{e^{1} - e^{2}}E_{(1)}^{e^{1} - e^{2}} \\&-  \sum_{J=1, \t{j} = 2}^{2,m}E^{(2)}_{e^{J} + e^{\t{j} }}F_{(2)}^{e^{J} + e^{\t{j} }} + F^{(2)}_{e^{J} + e^{\t{j} }}E_{(2)}^{e^{J} + e^{\t{j} }}
		- \sum_{J=1, \t{j} = 2}^{2,m}E^{(2)}_{e^{J} - e^{\t{j} }}F_{(2)}^{e^{J} - e^{\t{j} }} + F^{(2)}_{e^{J} - e^{\t{j} }}E_{(2)}^{e^{J} - e^{\t{j} }}\\ 
		&+ \sum_{J=1, j = 2}^{2,n}E^{(3)}_{e^{J} + e^{j}}F_{(3)}^{e^{J} + e^{j }} + F^{(3)}_{e^{J} + e^{j }}E_{(3)}^{e^{J} + e^{j }}
		+ \sum_{J=1, \t{j} = 2}^{2,n}E^{(3)}_{e^{J} - e^{j }}F_{(3)}^{e^{J} - e^{j }} + F^{(3)}_{e^{J} - e^{j }}E_{(3)}^{e^{J} - e^{j }}\\ 
		&+ \sum_{j < k = 2}^{n}E^{(4)}_{e^{j } + e^{k }}F_{(4)}^{e^{j } + e^{k }} + F^{(4)}_{e^{j } + e^{k }}E_{(5)}^{e^{j } + e^{k }} 
		+ \sum_{j < k = 2}^{n}E^{(4)}_{e^{j } - e^{k }}F_{(4)}^{e^{j } - e^{k }} + F^{(4)}_{e^{j } - e^{k }}E_{(4)}^{e^{j } - e^{k }}\\ &- \sum_{\t{j} < \t{k} = 2}^{m}E^{(5)}_{e^{\t{j} } + e^{\t{k} }}F_{(5)}^{e^{\t{j} } + e^{\t{k} }} + F^{(5)}_{e^{\t{j} } + e^{\t{k} }}E_{(5)}^{e^{\t{j} } + e^{\t{k} }} 
		- \sum_{\t{j} < \t{k} = 2}^{m}E^{(5)}_{e^{\t{j} } - e^{\t{k} }}F_{(5)}^{e^{\t{j} } - e^{\t{k} }} + F^{(5)}_{e^{\t{j} } - e^{\t{k} }}E_{(5)}^{e^{\t{j} } - e^{\t{k} }} \\&+ \sum_{j, \t{j} = 2}^{n,m}E^{(6)}_{e^{j} + e^{\t{j}} }F_{(6)}^{e^{j} + e^{\t{j}}} + F^{(6)}_{e^{j} + e^{\t{j}} }E_{(6)}^{e^{j} + e^{\t{j}}}+ \sum_{j, \t{j} = 2}^{n,m}E^{(6)}_{e^{j} - e^{\t{j}} }F_{(6)}^{e^{j}- e^{\t{j}}} + F^{(6)}_{e^{j} - e^{\t{j}} }E_{(6)}^{e^{j} - e^{\t{j}}}\\
		&= M_{A B}M^{A B} \;\; \forall A, B \in 1, \dots, 2n+2m
	\end{split}
\end{equation}
From here, we can lower the indices, and then use the commutation relations as well as the annihilation condition of $\Phi_{h}$ to write the action of the Casimir as
\begin{equation}
	\begin{split}
		\nabla \Phi_{h} &= \Big(\sum_{I = 1}^{2}\t{H}_{I}^{2} + \sum_{j=2}^{n}\t{H}_{j}^{2} +\sum_{\t{j}=2}^{m}\t{H}_{\t{j}}^{2} + \sum_{\pm}(\t{H}_{1} \pm \t{H}_{2} ) -  \sum_{\pm}\sum_{J=1, \t{j} = 2}^{2,m}(-\t{H}_{J} \pm \t{H}_{\t{j}})\\ &+\sum_{\pm}\sum_{J=1,j =2}^{2,n}(\t{H}_{J} \pm \t{H}_{j}) + \sum_{\pm}\sum_{k>j =2}^{n}(\t{H}_{j} \pm \t{H}_{k})\\ &+\sum_{\pm}\sum_{\t{j} < \t{k} = 2}^{m}(\t{H}_{\t{j}} \pm \t{H}_{\t{k}} ) - \sum_{\pm}\sum_{ j,\t{j} =2}^{n,m}(\t{H}_{j} \mp \t{H}_{\t{j} }  ) \Big)\Phi_{h}\label{casMIN}
	\end{split}
\end{equation}


\begin{thebibliography}{99}

 
\bibitem{Eden}
R.~J.~Eden, P.~V.~Landshoff, D.~I.~Olive and J.~C.~Polkinghorne,
``The analytic S-matrix,''

\bibitem{Elvang}
H.~Elvang and Y.~t.~Huang,
``Scattering Amplitudes,''
[arXiv:1308.1697 [hep-th]].

\bibitem{Strominger}
 A.~Atanasov, A.~Ball, W.~Melton, A.~M.~Raclariu and A.~Strominger,
 ``(2, 2) Scattering and the celestial torus,''
 JHEP \textbf{07}, 083 (2021)
 doi:10.1007/JHEP07(2021)083
 [arXiv:2101.09591 [hep-th]].

\bibitem{summary}
H.~Jiang
``Celestial OPEs and $ w_{1+\infty}$ algebra from worldsheet in string theory,''
[arXiv:2110.04255 [hep-th]].

V.~Chandrasekaran, E.~E.~Flanagan, I.~Shehzad and A.~J.~Speranza,
``Brown-York charges at null boundaries,''
[arXiv:2109.11567 [hep-th]].

L.~Donnay and R.~Ruzziconi,
``BMS Flux Algebra in Celestial Holography,''
[arXiv:2108.11969 [hep-th]].

S.~Banerjee, S.~Ghosh and P.~Paul,
``(Chiral) Virasoro invariance of the tree-level MHV graviton scattering amplitudes,''
[arXiv:2108.04262 [hep-th]].

S.~Pasterski,
``Lectures on Celestial Amplitudes''
[arXiv:2108.04801 [hep-th]].

P.~B.~Aneesh, G.~Comp\`ere, L.~P.~de Gioia, I.~Mol and B.~Swidler,
``Celestial Holography: Lectures on Asymptotic Symmetries,''
[arXiv:2109.00997 [hep-th]].

A.~M.~Raclariu,
``Lectures on Celestial Holography,''
[arXiv:2107.02075 [hep-th]].

M.~Campiglia and A.~Laddha,
``BMS Algebra, Double Soft Theorems, and All That,''
[arXiv:2106.14717 [hep-th]].

C.~Liu and D.~A.~Lowe,
``Conformal wave expansions for flat space amplitudes,''
JHEP \textbf{07} (2021), 102
doi:10.1007/JHEP07(2021)102
[arXiv:2105.01026 [hep-th]].


E.~Crawley, N.~Miller, S.~A.~Narayanan and A.~Strominger,
``State-operator correspondence in celestial conformal field theory,''
JHEP \textbf{09} (2021), 132
doi:10.1007/JHEP09(2021)132
[arXiv:2105.00331 [hep-th]].


W.~Fan, A.~Fotopoulos, S.~Stieberger, T.~R.~Taylor and B.~Zhu,
``Conformal blocks from celestial gluon amplitudes,''
JHEP \textbf{05}, 170 (2021)
doi:10.1007/JHEP05(2021)170
[arXiv:2103.04420 [hep-th]].

H.~Adami, D.~Grumiller, M.~M.~Sheikh-Jabbari, V.~Taghiloo, H.~Yavartanoo and C.~Zwikel,
``Null boundary phase space: slicings, news and memory,''
[arXiv:2110.04218 [hep-th]].

\bibitem{wikipedia}
 https://en.wikipedia.org/wiki/Anti-de\_Sitter\_space\#Definition\_and\_properties
 
 
\bibitem{Georgi}
H.~Georgi,
``Lie algebras in particle physics,''
Front. Phys. \textbf{54}, 1-320 (1999)

\end{thebibliography}
\end{document}